\documentclass{aa}  

\usepackage{graphicx}
\usepackage{txfonts}
\begin{document}

\title{Synthetic ngVLA line observations of a massive star-forming cloud}


\author{M. Juvela \inst{1},  E. Mannfors  \inst{1},  T. Liu\inst{2}, \and  L.V. T{\'o}th\inst{3}}

\institute{
Department of Physics, P.O.Box 64, FI-00014, University of Helsinki,
Finland, {\email mika.juvela@helsinki.fi}
\and
Shanghai Astronomical Observatory, Chinese Academy of Sciences, 
80 Nandan Road, Shanghai 200030, Peoples Republic of China
\and
Department of Astronomy, E{\"o}tv{\"o}s Lor{\'a}nd University, P{\'a}zmány P{\'e}ter s{\'e}t{\'a}ny 1/A,
H-1117 Budapest, Hungary
}

\date{Received September 15, 1996; accepted March 16, 1997}

 
\abstract
{Studies of the interstellar medium and the pre-stellar cloud evolution require spectral-line
observations that have high sensitivity and high angular and velocity resolution. Regions of high-mass
star formation are particularly challenging, because of line-of-sight confusion, inhomogeneous physical
conditions, and potentially very high optical depths.}
{We wish to quantify, to what accuracy the physical conditions within a massive star-forming cloud can
be determined from observations. We are particularly interested in the possibilities offered by the
Next Generation VLA (ngVLA) interferometer.}
{We use data from a magnetohydrodynamic simulation of star-formation in a high-density environment. We
concentrate on the study of a filamentary structure that has physical properties similar to a small
infrared-dark cloud. We produce synthetic observations for spectral lines observable with the ngVLA,
are analysed these to measure column density, gas temperature, and kinematics. Results are compared to
ideal line observations and the actual 3D model.}
{For a nominal cloud distance of 4\,kpc, ngVLA provides a resolution of $\sim$0.01\,pc even in its most
compact configuration. For abundant molecules, such as HCO$^+$, NH$_3$, N$_2$H$^+$, and CO isotopomers,
cloud kinematics and structure can be mapped down to sub-arcsec scales in just a few hours. For ${\rm
NH_3}$, a reliable column density map could be obtained for the entire $15\arcsec \times 40\arcsec$
cloud, even without the help of additional single-dish data, and kinetic temperatures are recovered to
a precision of $\sim$1\,K. At higher frequencies, the loss of large-scale emission becomes noticeable. 
The line observations are seen to accurately trace the cloud kinematics, except for the largest scales,
where some artefacts appear due to the filtering of low spatial frequencies. The line-of-sight
confusion complicates the interpretation of the kinematics, and the usefulness of collapse indicators
based on the expected blue asymmetry of optically thick lines is limited. }
{The ngVLA will be able to provide accurate data on the small-scale structure and the physical and
chemical state of star-forming clouds, even in high-mass star-forming regions at kiloparsec distances.
Complementary single-dish data are still essential for estimates of the total column density and the
large-scale kinematics.}
 
 
\maketitle
%

\section{Introduction}

Progress in star-formation (SF) studies is made via a combination of observations and numerical
simulations. Many open questions remain regarding the cloud formation, the balance between gravity,
turbulence, and magnetic fields at different scales, how filamentary structures are formed, how these
fragment, and how mass accretion takes place at different scales, from clouds to filaments and finally
to protostellar cores - and how all these processes differ between star-forming regions
\citep{Motte2018,Hacar2022,Pattle2022}. We also need better understanding of the interstellar medium
(ISM) itself, the gas and the dust, which are used as tracers of the SF process and affect, and are
affected by, the SF.

Understanding high-mass SF is particularly important as massive stars are crucial to many astrophysical
processes, and their ionising radiation and stellar winds, as well as the heavy elements produced by
supernovae affect their host galaxies \citep{Kennicutt2005}. Massive stars are formed partly in
infrared dark clouds (IRDCs), which are massive ($M\ga1000$\,M$_{\sun}$), cold (on average $T \sim
$10--20\,K) and dense ($\Sigma \sim 0.02$\,g\,cm$^{-2}$)
\citep{Peretto2010,Kainulainen2013,Tan2014,Lim2016}. 

Studies suggest that high-mass star-forming cores fragment as soon as they lose turbulent and magnetic
support \citep{Csengeri2011}, and numerical simulations show clouds not to be in equilibrium
\citep{Padoan2001,Vasquez2007}. Although simulations provide a direct handle on the dependencies
between SF and the environment where SF takes place, different views exist regarding the main causes of
especially the high-mass SF \citep{Zinnecker2007,Tan2018,Motte2018}.
It is still not clear whether high-mass stars form through competitive accretion
\citep{Larson1992,Bonnel2001,Bonnell2004} or core-accretion \citep{Larson1981,McKee1999,McKee2003} or
via large-scale processes that are driven mainly either by gravity
\citep{VazquezSemadeni2019,NaranjoRomero2022} or turbulent inertial flows
\citep{Padoan2020,Pelkonen2021} - or rather what is the relative importance of the different mechanisms
in different physical environments.

The SF theories are all supported by numerical simulations, although with some differences in the
assumed boundary conditions and included physics. It is essential that these paradigms are tested
against real observations in an objective way. Observations can be used to estimate physical parameters
(column densities, volume densities, temperatures), the kinematics and chemistry in different phases of
the SF process, and to quantify the resulting populations of clumps, filaments, and cores. These can
all be compared to the model predictions, using synthetic observations that take into account, at least
partially, the complexity of the source structure and the variations in the physical conditions along
the line of sight (LOS) and inside the finite telescope beam can be taken into account.

\begin{figure}
\sidecaption
\centering
\includegraphics[width=9cm]{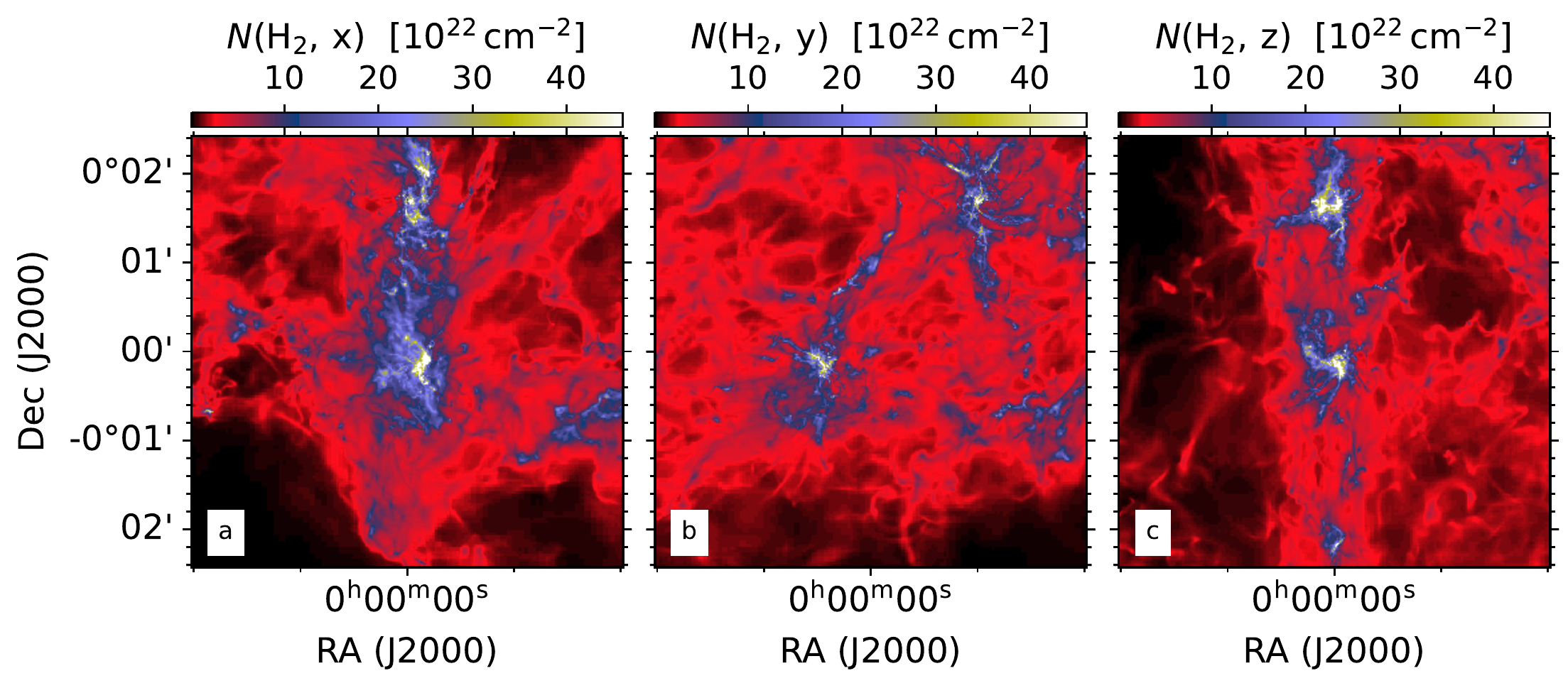}
\caption{
Column density of the selected cloud region viewed from three orthogonal directions. Frame a
corresponds to the direction that is selected for the analysis of synthetic observations. The actual
3D spatial separation between the northern and southern density peaks is shown best in frame b (structures
in the upper right and lower left part of the plot).
}
\label{fig:colden}
\end{figure}

\begin{figure}
\centering
\sidecaption
\includegraphics[width=9cm]{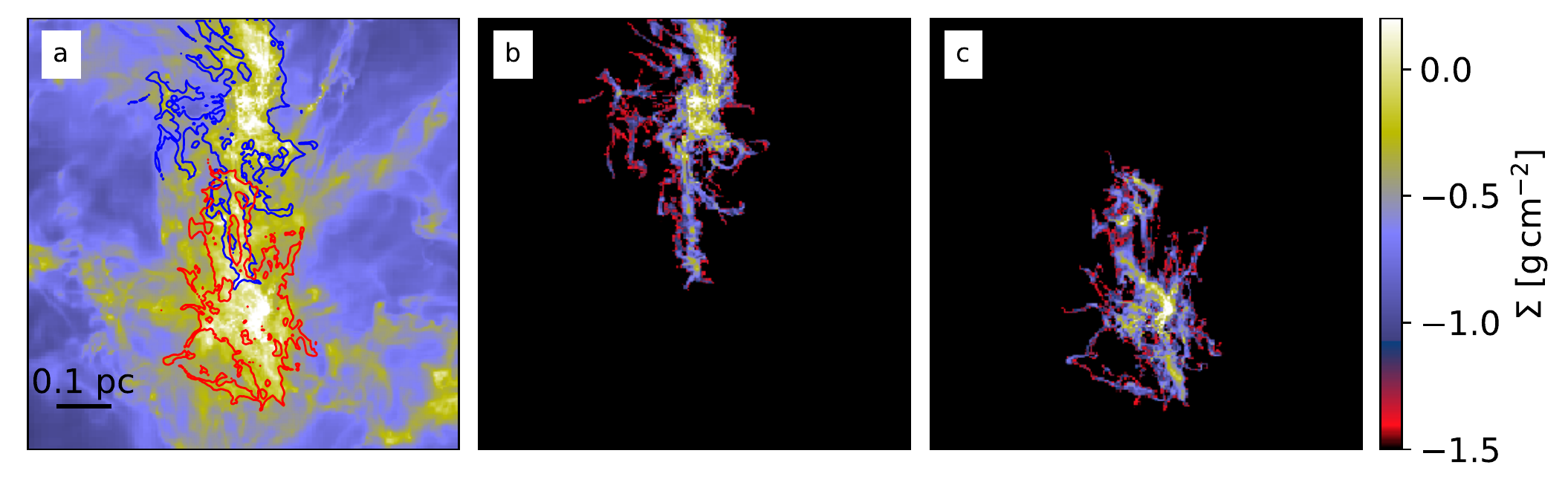}
\caption{
Surface mass density of the whole model cloud (frame a) and separately for the northern and southern
clumps (frames b and c). The clumps are defined using a density isosurface at $2\times 10^5\,{\rm
cm^{-3}}$, and their outlines are overplotted also in frame a.
}
\label{fig:mass_centres}
\end{figure}

To accurately test star-formation theories, one needs observations that cover large areas with high
sensitivity and fidelity and, on the other hand, have the resolution to probe the small-scale core
fragmentation and the initial stages of protostellar collapse. Recent continuum observations, for
example the surveys conducted with the Herschel Space Observatory \citep{Pilbratt2010}, have resulted
in significant advances in the understanding of the structure of the star-forming clouds and the
large-scale context of SF. However, spectral line observations remain crucial, to mitigate the
effects of LOS confusion and to gain direct access to the physical properties, kinematics, and
chemistry of the gas component. 

Radio interferometers are essential for studies of the small structures, the cloud cores with sizes
below $\sim$0.1\,pc or even further the core fragmentation down to $\sim$1000\,au scales
\citep{Tokuda2020,Sahu2021}. Interferometry is also needed for the more distant targets, including the
typical high-mass star-forming clouds that are at kiloparsec distances. 
\citep{Liu2020_ATOMS-I,Beuther2021,Oneill2021,Motte2022_IMF_I}. Instruments like the Atacama Large
Millimeter/submillimeter Array (ALMA\footnote{https://www.almaobservatory.org}, the Submillimetre Array
(SMA\footnote{https://lweb.cfa.harvard.edu/sma/}), and the IRAM NOrthern Extended Millimeter Array
(NOEMA\footnote{https://www.iram-institute.org/EN/noema-project.php}) already provide sub-arcsecond
resolution.

Next-generation Very Large Array (ngVLA\footnote{https://ngvla.nrao.edu}) is the planned extension of
the Karl G. Jansky Very Large Array (VLA\footnote{https://www.vla.nrao.edu}). It will cover frequencies
1.2-116\,GHz with (at the highest frequencies) a maximum instantaneous bandwidth of 20\,GHz and with a
sensitivity and angular resolution potentially exceeding those of both the current VLA and ALMA
instruments. The ngVLA will also complement the radio-frequency coverage of ALMA and
SKA\footnote{https://www.skatelescope.org} that, respectively, operate mainly at higher and lower
frequencies. It will be complementary also by providing high-resolution observations of the northern
sky. The ngVLA will be able to observe basic transitions of several key molecules, such as NH$_3$,
some of which are not accessible to ALMA but are important for SF studies and especially for the study
of the cold ISM and the early phases of the SF process. 

In this paper we study, as a test case, synthetic observations of a massive filamentary cloud. The
target has properties similar to a infrared dark cloud (IRDC) capable of giving birth to high-mass
stars. The cloud model is obtained from a magnetohydrodynamic (MHD) simulation, which is post-processed
with radiative transfer calculations to predict line emission in several transitions that are
accessible to the ngVLA. The simulated line data are processed with the CASA
program\footnote{https://casa.nrao.edu} to make predictions for actual ngVLA observations, including
realistic noise and effects of the interferometric mode of observations. 
This is part of an ngVLA community study, where we investigate the use of the ngVLA interferometer for
observations of star-forming clouds. In the present paper, the emphasis is on moderately extended
structures, and we concentrate on observations with the most compact ngVLA antenna configuration.
Smaller scales (e.g. sub-fragmentation of cores) will be addressed in future publications.

The contents of the paper are the following. Section~\ref{sect:simulations} describes the MHD runs
(Sect.~\ref{sect:MHD}), radiative transfer modelling (Sect.~\ref{sect:RT}), and CASA simulations
(Sect.~\ref{sect:CASA}). The synthetic observations are analysed in Sect.~\ref{sect:results}, regarding
column densities (Sect.~\ref{sect:colden}) and temperatures (Sect.~\ref{sect:temperature}) and cloud
kinematics (\ref{sect:kinematics}). We discuss the results in Sect.~\ref{sect:discussion} before
listing the final conclusions in Sect.~\ref{sect:conclusions}.

\section{Simulated observations} \label{sect:simulations}

\subsection{MHD simulations}  \label{sect:MHD}

As a starting point for the synthetic observations, we use the density and velocity fields from the MHD
simulation described in \citet{Haugbolle2018}. The MHD run covers a volume of (4\,pc)$^3$, using octree
spatial discretisation with a root grid of 256$^3$ cells and up to five levels of refinement, reaching
a best linear resolution of 100\,au ($4.88 \times 10^{-4}$\,pc). The refinement was based on density,
especially to resolve the collapsing cores. Sink particles were used to represent collapsed regions
where the density exceeded the average density by a factor of 10$^5$-10$^6$ and where stars are thus
forming. The total mass contained in the (4\,pc)$^3$ box is about 3000\,M$_{\sun}$. In the snapshot
used in this paper, over 400 stars have already formed, and these are also used in the subsequent
radiative transfer modelling as radiation sources. Details of the MHD simulation and a study of formed
stellar populations can be found in \citet{Haugbolle2018}.

\subsection{Radiative transfer modelling} \label{sect:RT}

We concentrate on the densest part of the model and extracted a (1.84\,pc)$^3$ subvolume out of the
full (4\,pc)$^3$ MHD run. The subvolume contains 230 young stars with masses ranging from sub-solar to
36\,$M_{\sun}$ and includes two main density enhancement that, in the selected view direction ``x'',
appear as a single elongated structure (Fig.~\ref{fig:colden}a). We refer to this structure as ``the
filament''. In 3D, it consists of a northern and a southern clumps, which are connected also in 3D, but
only by narrow bridges that are inclined by about 45 degrees with respect to the observer LOS
(Fig.~\ref{fig:colden}b). The 3D nature of the object is therefore different from its appearance in the
projected images. Figure~\ref{fig:mass_centres} shows the mass surface density towards the direction
``x''. The figure also shows the outlines of the northern and southern clumps, which are in 3D defined
by density isosurfaces.

We started by modelling the dust component, using the dust properties from
\citet{Weingartner_Draine_2001}, the Case B that has a selective extinction of $R_{\rm V}=5.5$, the
high value being appropriate for dense clouds. The calculations were carried out with the radiative
transfer programme SOC \citep{Juvela_2019_SOC}, using the full octree discretisation of the MHD model.
The dust heating included an isotropic external field \citep{Mathis1983} and the contribution of the
embedded stars, which were modelled as as blackbody point sources, with the given luminosities and
effective temperatures \citep{Haugbolle2018}. The radiative transfer calculations solved the dust
temperature $T_{\rm dust}$ for each model cell, assuming that the grains are at equilibrium with the
radiation field. The dust-temperature distribution has its mode at 14.3\,K and extends below 10\,K in
dense, non-protostellar cores. The temperature structure is not fully resolved in the immediate
vicinity of stars, but temperatures at and above 100\,K are reached in nearly 0.1\% of the cells.
Because the spatial grid is highly refined close to the star-forming cores, this corresponds to a much
smaller fraction of the cloud volume. The analysis of the synthetic observations of dust emission (and
polarisation as a proxy of the magnetic fields) are deferred to a future publication. In this paper,
only the information of the estimated large-grain dust temperatures is used. Appendix~\ref{sect:Tdust}
shows a colour-temperature map, how the dust temperature distribution would appear based on
160-500\,$\mu$m continuum observations.

The line modelling is based on the density and velocity fields of the MHD simulation and the results of
the dust modelling, assuming that the dust temperature serves as a proxy for the gas kinetic
temperature, $T_{\rm kin}\approx T_{\rm dust}$. The line modelling was carried out with the radiative
transfer program LOC \citep{Juvela_2020_LOC}, with the molecular data obtained from the LAMDA database
\citep{Schoier2005}. The calculations solve the non-LTE excitation of the molecules in each cell and
provide spectral line maps towards chosen directions. Spectra were computed for $^{13}$CO(1-0),
C$^{18}$O(1-0), HCO$^{+}$(1-0), ${\rm H^{13}CO^{+}}$(1-0), ${\rm N_2 H^{+}}$(1-0), ${\rm NH_3}$(1,1),
and ${\rm NH_3}$(2,2) lines. In the case of ${\rm N_2 H^{+}}$(1-0) and ${\rm NH_3}$(1,1), the hyperfine
structure was taken into account by assuming LTE conditions between the hyperfine components
\citep{Keto1990}. 

The assumed peak fractional abundances are 
$2\times 10^{-6}$ for $^{13}$CO,
$3\times 10^{-7}$ for ${\rm C^{18}O}$, 
$5\times 10^{-10}$ for HCO$^{+}$, 
$1\times 10^{-11}$ for ${\rm H^{13}CO^{+}}$, 
$3\times 10^{-8}$ for ${\rm NH_3}$, and 
$1\times 10^{-9}$ for ${\rm N_2 H^{+}}$.
The abundances are further scaled with an additional density dependence $n({\rm H}_2)^{2.45}/(3.0\times
10^8+n({\rm H}_2)^{2.45})$ \citep[cf.][]{Glover2010}. This decreases the abundances at densities below
$n({\rm H}_2)=10^{4}$\,cm$^{-3}$, reduces them by a factor of ten at $n({\rm H}_2)=10^{3}$\,cm$^{-3}$
and makes them negligible at the lowest densities, where the gas can be assumed to be mostly atomic. We
do not model the depletion that could decrease the abundances at the highest densities, especially
within the pre-stellar cores.

The assumption $T_{\rm kin}\approx T_{\rm dust}$ is accurate at densities close to and above
$10^5$\,cm$^{-3}$, where the gas-dust collisions bring the gas kinetic temperature to within a few
degrees of the dust temperature \citep{Goldsmith2001, Young2004, Juvela2011}. For example for ${\rm
N_{2}H^{+}}$(1-0), most emission comes from such high-density gas, because the critical density of the
transition is close to $n({\rm H}_2) = 10^5$\,cm$^{-3}$. There is a range of densities around
$n({\rm H}_2) \approx 10^4$\,cm$^{-3}$ where we may underestimate the kinetic temperature and the ${\rm
N_{2}H^{+}}$ abundance is not negligible. However, the effect of this uncertainty is smaller than that
of the assumed absolute abundances. HCO$^{+}$ has a similarly high critical density. The $T_{\rm
kin}\approx T_{\rm dust}$ approximation is less appropriate for ${\rm C^{18}O}$ and especially for
$^{13}$CO, because its higher optical depth leads to significant excitation at lower densities.
However, in the absence of local heating sources, $T_{\rm dust}$ remains close to 20\,K in the less
dense regions, and if the gas is assumed to be in molecular form (the lack of photodissociation also
implying reduced heating via the photoelectric effect), the gas kinetic temperatures should not be much
higher.

The fractional abundances are generally subject to large uncertainty.  The typical $^{13}$CO abundance
is $\sim 10^{-6}$ \citep[e.g.][]{Dickman1978, Roueff2021}, but the [$^{13}$CO]/[${\rm C^{18}O}$] ratio
can vary significantly around the cosmic abundance ratio $\sim$5.5 \citep{Myers1983}. Values $3-10$
have been reported in dark clouds and high-mass star-forming regions alike \citep{Paron2018, Areal2018,
Roueff2021}. In our simulations, the  ratio is [$^{13}$CO]/[${\rm C^{18}O}$]=6.67, which is
close to the average values found in IRDCs \citep{Du2008, Areal2019}. 
\citet{Pirogov1995} reported an average value [HCO$^{+}$]/[${\rm H^{13}CO^{+}}$]=29 for a sample of
dense clouds. \citet{RodriguezBaras2021} examined nearby clouds and, for example in Orion, found
roughly similar values. However, the distribution of estimates is very wide and extends up to values
similar to the isotopic ratio $^{12}$C/$^{13}$C$\sim$90 in the solar neighbourhood \citep{Milam2005}.
We use a ratio [HCO$^{+}$]/[${\rm H^{13}CO^{+}}$]=50. The assumed abundance [HCO$^{+}$]=$5 \times
10^{-10}$ is close to the lower limit found in the IRDC survey of \citet{Vasyinina2011} or in
\citet{RodriguezBaras2021}. For most clouds, the estimated abundances are higher, by up to a factor of
ten, and our simulations are therefore somewhat pessimistic regarding the HCO$^{+}$ and ${\rm
H^{13}CO^{+}}$ line intensities \citep[see also][]{Blake1987,Sanhueza2012}.
For ammonia, the selected value [${\rm NH_3}$]=10$^{-8}$ is typical of IRDCs \citep{Chira2013,
Sokolov2017}.
Finally, the estimates of [${\rm N_{2}H^{+}}$] vary more than one order of magnitude even in samples of
dense clouds. \citet{Ryabukhina2021} found in the IRDC G351.78-0.54 values [${\rm
N_{2}H^{+}}$]=0.5-2.5$\times10^{-10}$, which are similar to the values 1-4$\times 10^{-10}$ reported
for low-mass cores \citep{Caselli2002a}. However, in \citet{Gerner2014}, the values tended to be
slightly below [${\rm N_{2}H^{+}}$]=$10^{-9}$ for IRDCs and slightly higher for high-mass protostellar
objects. Using data from the MALT90 survey \citep{Foster2011}, \citet{Miettinen2014} found an average
of [${\rm N_{2}H^{+}}$]=$1.6\times 10^{-9}$, both infrared-dark and infrared-bright sources exhibiting
similar abundances. Our assumed value [${\rm N_{2}H^{+}}$]=$1\times 10^{-9}$ is therefore closer to
these higher estimates.

In the present paper, our main interest is in the comparison between ``a model'' and the synthetic
observations made of that particular model. A high degree of physical accuracy in the underlying cloud
description is therefore less crucial, and this applies both to the assumed temperature and abundance
distributions.

The radiative transfer modelling resulted in spectral-line maps that cover the model with a pixel size
that was set equal to the smallest cell size in the 3D model. For the assumed cloud distance of 4\,kpc,
one pixel (equal to 100\,au or $4.88\times 10^{4}$\,pc) corresponds to an angular size of
0.025$\arcsec$. The velocity resolution of the extracted spectra was set to $\sim$0.1\,km\,s$^{-1}$,
which is of the order of the thermal line broadening and much below the total observed line widths that
are typically $\sim$1\,km\,s$^{-1}$ or larger. 

The spectra that are obtained directly from the radiative transfer calculations are in the following
called the ``ideal'' observations. The radiative transfer program LOC is not based on Monte Carlo
method, and the spectra do not therefore contain random Monte Carlo noise. The errors are only due to
the sampling of the radiation field with a finite number of rays. This could have a small effect when
line data are compared directly to the properties of the 3D model, but it does not directly affect the
comparison between the ``ideal'' observations and the synthetic ngVLA observations, because the latter
are based on the former.

\begin{figure}
\centering
\includegraphics[width=9cm]{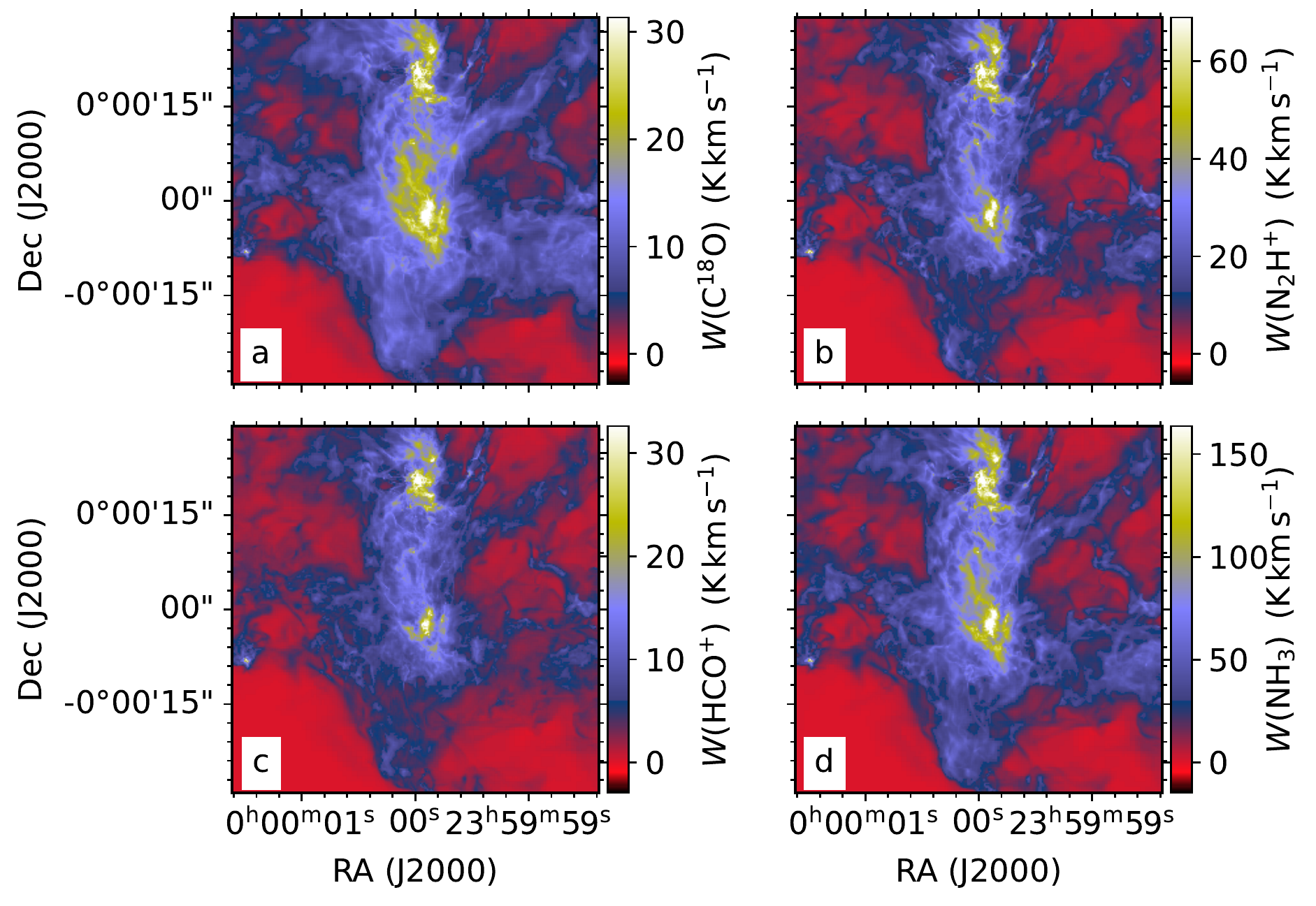}
\caption{
Comparison of integrated intensities $W$ in ideal ${\rm C^{18}O}$, ${\rm N_2 H^{+}}$, HCO$^{+}$,
and ${\rm NH_3}$ observations.
}
\label{fig:W}
\end{figure}

\subsection{CASA simulations} \label{sect:CASA}

The spectral cubes produced by the radiative transfer calculations were converted to synthetic ngVLA
observations with the help of the CASA
simulator\footnote{https://casaguides.nrao.edu/index.php/Simulating\_ngVLA\_Data-CASA5.4.1}.  Because
our emphasis is on extended rather than sub-core-scale structure, we concentrate on observations made
with the Core Subarray, which consists of 94 antennas with maximum baselines of 1.3\,km. 

The ngVLA observation were simulated with the CASA {\rm simobserve} script, adding noise according to
the published ngVLA characteristics\footnote{https://ngvla.nrao.edu/page/performance}. The simulated
raw observations were also processed with the CASA program, where maps were made with the {\tt tclean}
algorithm, with robust weighting (robust=0.5) and automatic multi-threshold masking. 

The target field was set at zero declination, which results in small ellipticity in the synthesised
beam. For the Core antenna configuration, after adding a small amount of tapering, the resolution
ranges from $2.6\arcsec \times 2.2\arcsec$ for the ammonia lines near 23.7\,GHz to about $0.58\arcsec
\times 0.50\arcsec$ for the C$^{18}$O and $^{13}$CO lines near 110\,GHz.  We used two pointings that
are separate by 14$\arcsec$ in the latitude. The corresponding map sizes range from nearly circular
ammonia maps, with a diameter of 3.6$\arcmin$, to more elliptical coverage of $0.87\arcmin\times
0.77\arcmin$ at the highest frequencies. This is sufficient to cover the cloud filament that has a
length less than $\sim$0.7$\arcmin$ (0.8\,pc at the distance of 4\,kpc). The nominal simulations
correspond to an observing time of six hours with the Core antenna configuration. All observations are
corrected for the main beam.

\section{Results}  \label{sect:results}

\subsection{Line maps} \label{sect:maps}

Figure~\ref{fig:W} shows maps of integrated intensity for the ideal observations (the direct output
from the radiative transfer modelling without beam convolution, observational noise, or the effects of
interferometric observations). Apart from differences in the signal level, the molecules (${\rm
C^{18}O}$, ${\rm N_2 H^{+}}$, HCO$^{+}$, and ${\rm NH_3}$) show only small differences that result from
different optical depths and critical densities. The ${\rm N_2 H^{+}}$ and ${\rm NH_3}$ integrated
intensities include the emission from all hyperfine components.

\begin{figure}
\centering
\includegraphics[width=9cm]{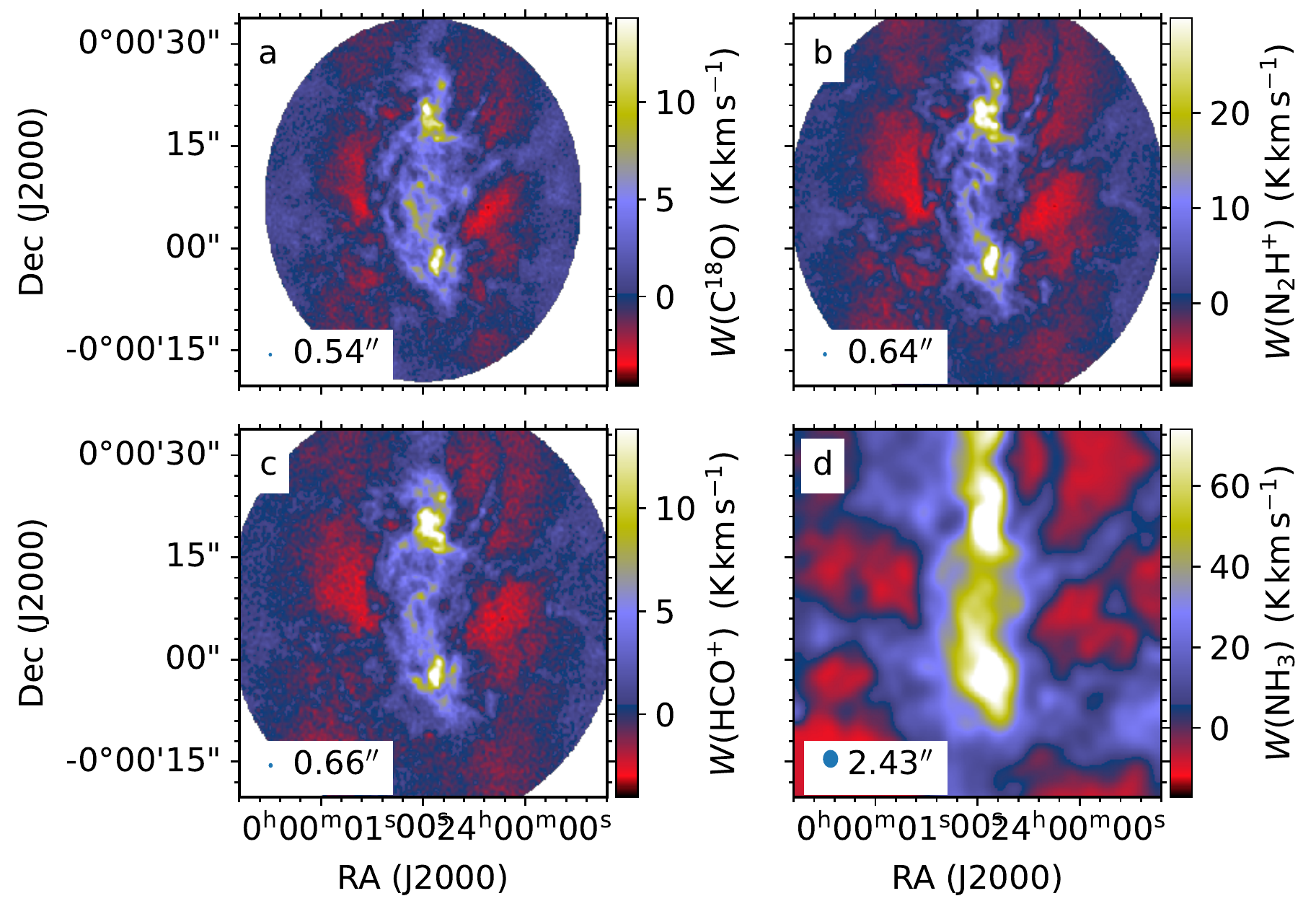}
\caption{
Comparison of integrated intensity $W$ in simulated ngVLA observations (Core antenna configuration with
six hours of observation). As in Fig.~\ref{fig:W}, the frames show the data for ${\rm C^{18}O}$, ${\rm
N_2 H^{+}}$, HCO$^{+}$, and ${\rm NH_3}$.
}
\label{fig:Wobs}
\end{figure}

\begin{figure}
\centering
\includegraphics[width=9cm]{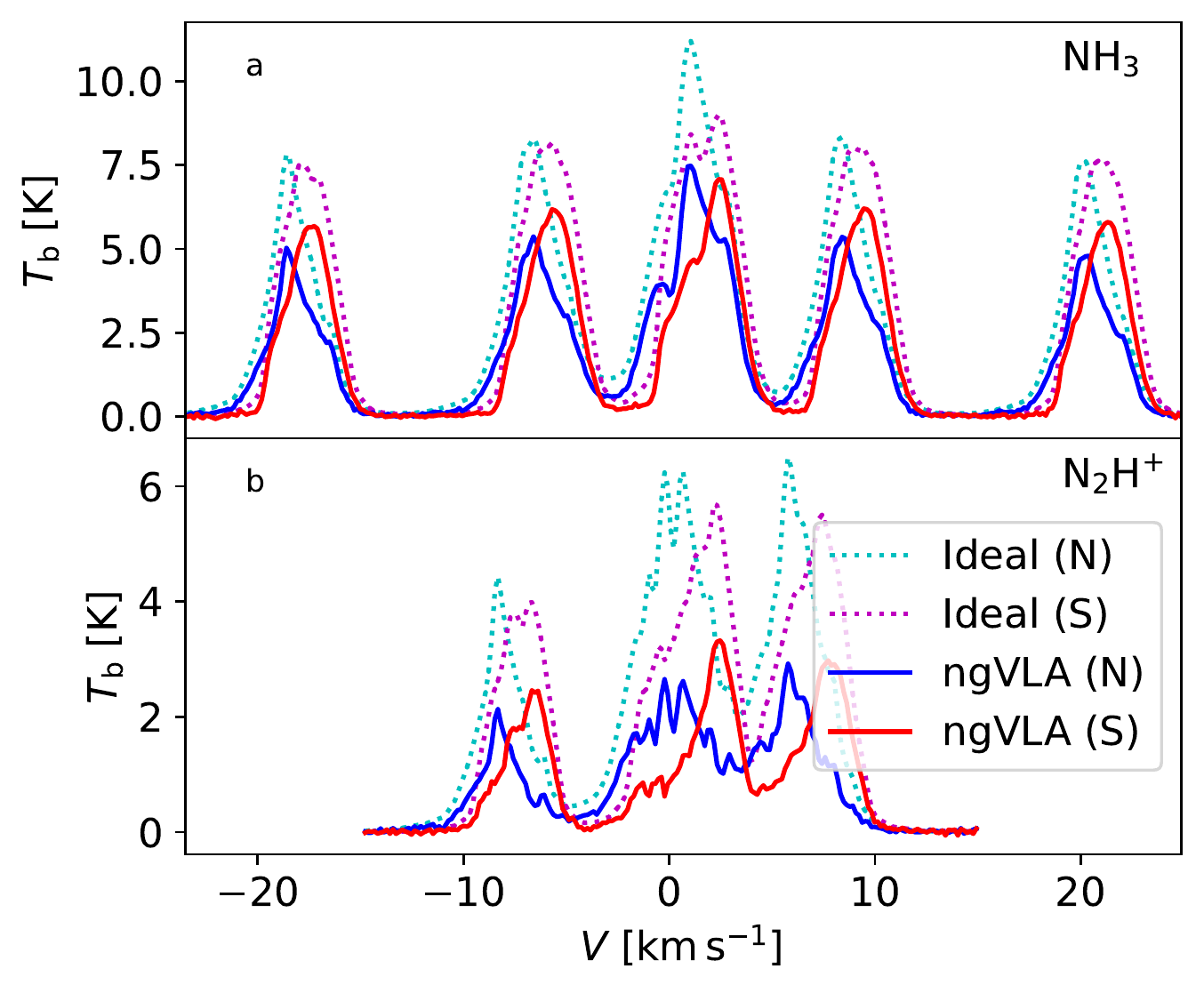}
\caption{
Average ideal and ngVLA-observed ${\rm NH_3}$ and ${\rm N_2 H^{+}}$ spectra towards the northern (N)
and the southern (S) clumps. The dotted lines correspond to ideal observations and the solid lines to
the nominal synthetic ngVLA observations.
}
\label{fig:plot_spectra}
\end{figure}

Figure~\ref{fig:Wobs} is the corresponding plot for the nominal synthetic ngVLA observations (six hours
with the Core array). The noise is sufficiently low not to be clearly visible in the integrated
intensity. The main differences compared to Fig.~\ref{fig:W} are the loss of some extended emission,
for example in the South-West part of the ${\rm C^{18}O}$ map, which is also visible as lower
integrated intensity. There are some regions of negative signal around of the filament. These artefacts
might be reduced by more careful manual data reduction and especially by the inclusion of single-dish
data. However, at small scales, the correspondence to the ideal maps is good and mainly limited by the
beam size of the synthetic observations. Figure~\ref{fig:plot_spectra} shows ${\rm NH_3}$ and ${\rm N_2
H^{+}}$ spectra that are averaged over parts of the northern and southern clumps where the maximum LOS
density exceeds $2\times 10^6$\,cm$^{-3}$ (about half of the areas shown in
Fig.~\ref{fig:mass_centres}). In spite of the complex 3D structure of the target areas (see
Appendix~\ref{app:LOS_emission}), the average spectra are not far from Gaussian.

\begin{figure}
\sidecaption
\centering
\includegraphics[width=8.8cm]{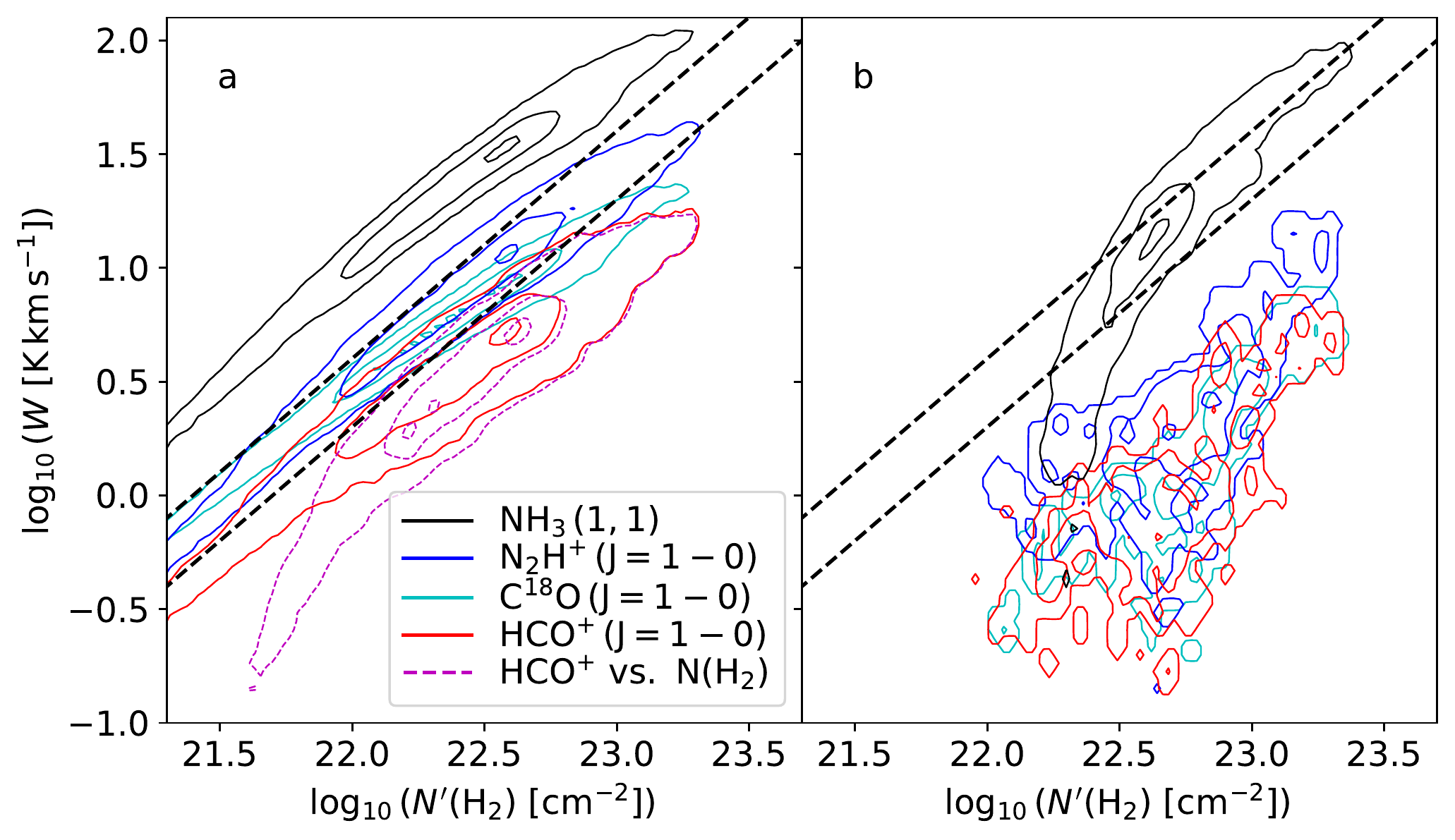}
\caption{ 
Integrated intensities of selected molecules as function of $N^{\prime}({\rm H}_2)$, which is the true
model column density that has been scaled to take into account the density dependence of molecular
abundances. Frame a shows the data for ideal observations (full model resolution) and frame b for the
simulated ngVLA observations (resolution of ngVLA data). The contours are drawn at 10\%, 50\%, and 90\%
of the peak point density (pixels per logarithmic column-density and line-area interval). For HCO$^{+}$
in frame a, the distribution is plotted also against $N({\rm H}_2)$, showing the effect of lower
average abundances at lower densities.
For reference, the dashed black lines show two linear relationships that are separated by a factor of
two along the $W$ axis:
$W=  2   \cdot 10^{-22} \cdot N^{\prime}({\rm H}_2)\,{\rm K\, km \,s^{-1}}$ and 
$W=  4   \cdot 10^{-22} \cdot N^{\prime}({\rm H}_2)\,{\rm K\, km \,s^{-1}}$.
}
\label{fig:W_vs_true}
\end{figure}

Figure~\ref{fig:W_vs_true} shows the observed integrated line intensities as a function of column
density. The x-axis is here a modified value $N^{\prime}({\rm H}_2)$, which is the true model column
density scaled with the density dependence that was used in setting the molecular abundances. The
density dependence was the same for all molecules. Thus, the actual column density of each molecule is
equal to $N^{\prime}({\rm H}_2)$ multiplied with the peak abundances of the molecule
(Sect.~\ref{sect:RT}). In an ideal case, the relationship between $W$ and $N^{\prime}({\rm H}_2)$ should
be linear.

The first frame of Fig.~\ref{fig:W_vs_true} shows the results for ideal observations. There are minor
deviations from linear relationships, with only small effects from temperature (and
excitation-temperature) variations and some saturation at the highest optical depths. The same factors
affect the dispersion, which is of the order of a factor of two in the direction of the $W$ axis.
Figure~\ref{fig:W_vs_true}b shows the same for the nominal synthetic ngVLA observations (without
single-dish data). The observed $W$ is plotted against the same modified column density
$N^{\prime}({\rm H}_2)$ as above, but convolved to the angular resolution of the observations. The
correlation between column density and integrated intensity $W$ remains fair at the highest column
densities, but the observed $W$ values are now lower because of interferometric filtering. The ammonia
observations benefit from the lower frequency and the correspondingly larger synthesised beam (lower
noise) and larger main beam (larger maximum recoverable scale). 

If the line area were plotted against the true column density $N({\rm H}_2)$ rather than against
$N^{\prime}({\rm H}_2)$, the $W$ values at the low-column-density part of the plot would appear to
drop, although the actual change is in the x-axis variable.  The values are lower because the average
fractional abundances decrease towards lower column densities. In our case, this amounts to more than a
factor of three reduction in the $W$ values at $N({H}_2)\sim 5\cdot 10^{21}$\,cm$^{-2}$, compared to
the value seen at $N^{\prime}({H}_2))\sim 5\cdot 10^{21}$\,cm$^{-2}$.  However, this difference is
entirely dependent on the ad hoc assumption of the fractional abundances.

\subsection{Column densities} \label{sect:colden}

Column densities can be estimated with different spectral lines or line combinations 
(Appendix~\ref{app:colden}). We examine four cases: (1) assuming optically thin ${\rm C^{18}O}$ lines
with $T_{\rm ex}$=15\,K, (2) using a combination of ${\rm H^{13}CO^{+}}$ and HCO$^{+}$ lines, or using
the hyperfine structure of either (3) the ${\rm N_2 H^{+}}$(1-0) or (4) the ${\rm NH_3}$(1, 1) lines.
All estimates are converted to $N({\rm H}_2)$ by using the maximum fractional abundance in the model
cloud. To eliminate the effect of abundance variations, the true model column densities are also
rescaled to take into account the LOS abundance variations, and, once these are convolved to the
resolution of the observations, there should ideally be a one-to-one correspondence to the values
derived from the synthetic observations. 

We fit the spectra with one or more Gaussian components or, in the case of ${\rm N_2 H^{+}}$ and ${\rm
NH_3}$, with the hyperfine structure and one velocity component. Appendix~\ref{sect:spectral_fits}
shows examples of ${\rm C^{18}O}$ spectra observed towards the northern core that are fitted with
different number of Gaussian components. Both the northern and southern cores show a very complex
velocity structure, but the relevant line parameters can still be mostly approximated using the single
Gaussian. Towards the centre of the field, the emission from the two LOS regions overlaps, and the
spectra show two equally strong velocity components that are separated by some 2\,km\,s$^{-1}$ in
velocity.

\begin{figure*}
\centering
\sidecaption
\includegraphics[width=12cm]{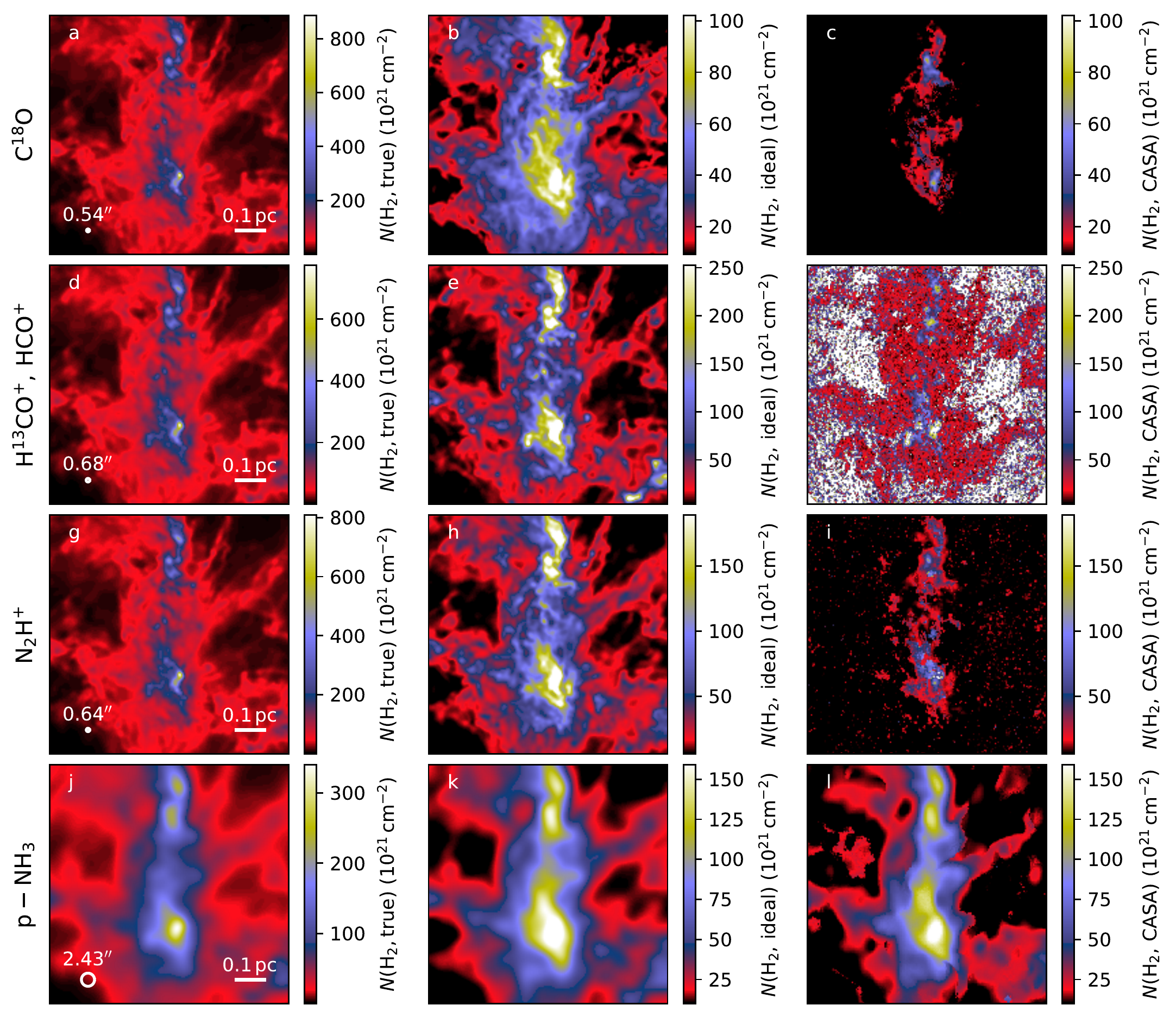}
\caption{
Comparison of true and estimated column densities. The three columns correspond to the true column
density, the column density estimated from ideal observations, and the column density derived from the
nominal synthetic ngVLA observations. Results are shown for C$^{18}$O (first row; optically-thin
approximation), combination of ${\rm H^{13}CO^{+}}$ and ${\rm HCO^{+}}$ lines (second row), ${\rm
N_2H^{+}}$ (third row; fit of hyperfine structure), and NH$_3$ (bottom row; fit of hyperfine
structure). Observational estimates have been scaled to $N({\rm H}_2)$ using the maximum
fractional-abundance value in the modelling. In the first column, the true column densities are also
correspondingly scaled down according to the spatial variation of the relative abundances.
}
\label{fig:colden_maps}
\end{figure*}

In Fig.~\ref{fig:colden_maps}, the first column shows the true column density that is modified
according to the common density dependence of the abundances. The second column shows the corresponding
estimates derived from ideal observations, and the third column the estimates based on the synthetic
ngVLA observations. The analysis is based on Gaussian fits or, in the case of ${\rm N_2 H^{+}}$ and
${\rm NH_3}$ lines, hyperfine fits, all with a single fitted velocity component. There are already
significant differences between the first two columns that are caused by the assumptions of the
column-density estimation: LOS homogeneity, the fixed $T_{\rm ex}$ value in the case of ${\rm
C^{18}O}$, and the inaccuracy of representing the spectra with single Gaussian velocity components.

\begin{figure}
\centering
\sidecaption
\includegraphics[width=9cm]{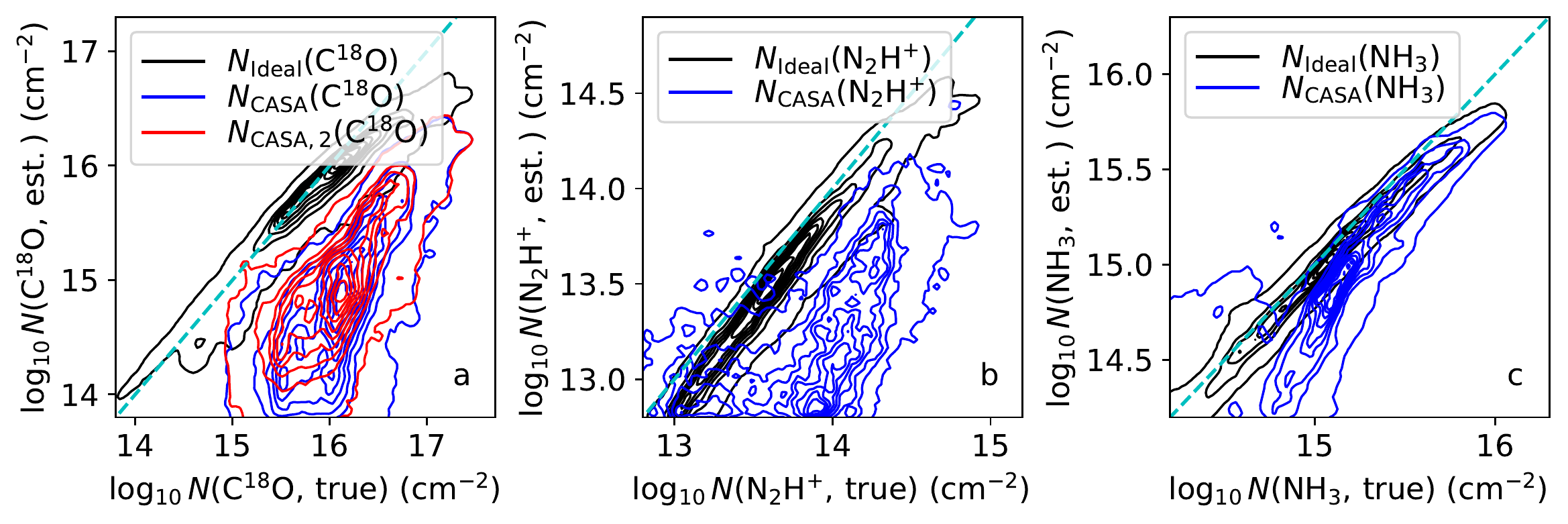}
\caption{
Correlations between column density estimates and the true column density. 
The y-axis shows the ideal and simulated ngVLA estimates, and the x-axis the true model column
density of the molecule in question.
Results are shown for ${\rm
C^{18}O}$, ${\rm N_2 H^{+}}$, and ${\rm NH_3}$. These correspond to the maps in Fig.~\ref{fig:colden},
with the addition of ${\rm C^{18}O}$ estimates based on the fitting of two Gaussian velocity components
($N_{\rm CASA, 2}$ in frame a). There are nine linearly-spaced contour levels, up to the maximum point
density (pixels per logarithmic intervals of true and observed column density).
}
\label{fig:plot_colden_cor}
\end{figure}

Figure~\ref{fig:plot_colden_cor} plots the column-density estimates from ${\rm C^{18}O}$, ${\rm N_2
H^{+}}$, and ${\rm NH_3}$ against the true column density. The plots include estimates from both
the ideal observations and the synthetic ngVLA observations. The ideal ${\rm C^{18}O}$ observations
provide accurate values, the adopted $T_{\rm kin}=15$\,K not resulting in a significant offset. Bias
appears only at high column densities, $N({\rm C^{18}O})>10^{16}$\,cm$^{-2}$, where the lines are no
longer optically thin or possibly because the $T_{\rm kin}$ values of the model cloud are
systematically lower in regions of high density. In contrast, the ngVLA estimates are clearly lower,
mirroring the trends in Fig.~\ref{fig:W_vs_true}. Gaussian fits with two velocity components usually
result in only small changes, except for some low-column-density regions.

For ${\rm N_2 H^{+}}$, synthetic ngVLA observations reach high S/N only in the high-density part of the
filament (Fig.~\ref{fig:colden_maps}i) and the estimates are below the true values. For ${\rm NH_3}$,
the beam size is larger (2.4$\arcsec$ compared to 0.64$\arcsec$ for ${\rm N_2 H^{+}}$), the S/N (and
the main-beam size) is similarly larger, and the ngVLA observations provide more accurate estimates up
to the highest column densities. At ammonia column densities $\sim 10^{15}$\,cm$^{-2}$ and below, the
estimates fall below the true values. This is probably due to the filtering of large-scale emission,
although the assumption of a single velocity component may also bias the results in some regions.

\subsection{Kinetic temperature} \label{sect:temperature}

We estimated the gas kinetic temperature with the combination of ${\rm NH_3}$(1,1) and ${\rm
NH_3}$(2,2) lines (Appendix~\ref{app:colden}). The analysis assumes a homogeneous and isothermal
medium, but the observations should give a good approximation for the mass-weighted temperature, as
long as the LOS variations and the optical depths are not extreme \citep{Juvela2012}.

\begin{figure}
\centering
\includegraphics[width=9cm]{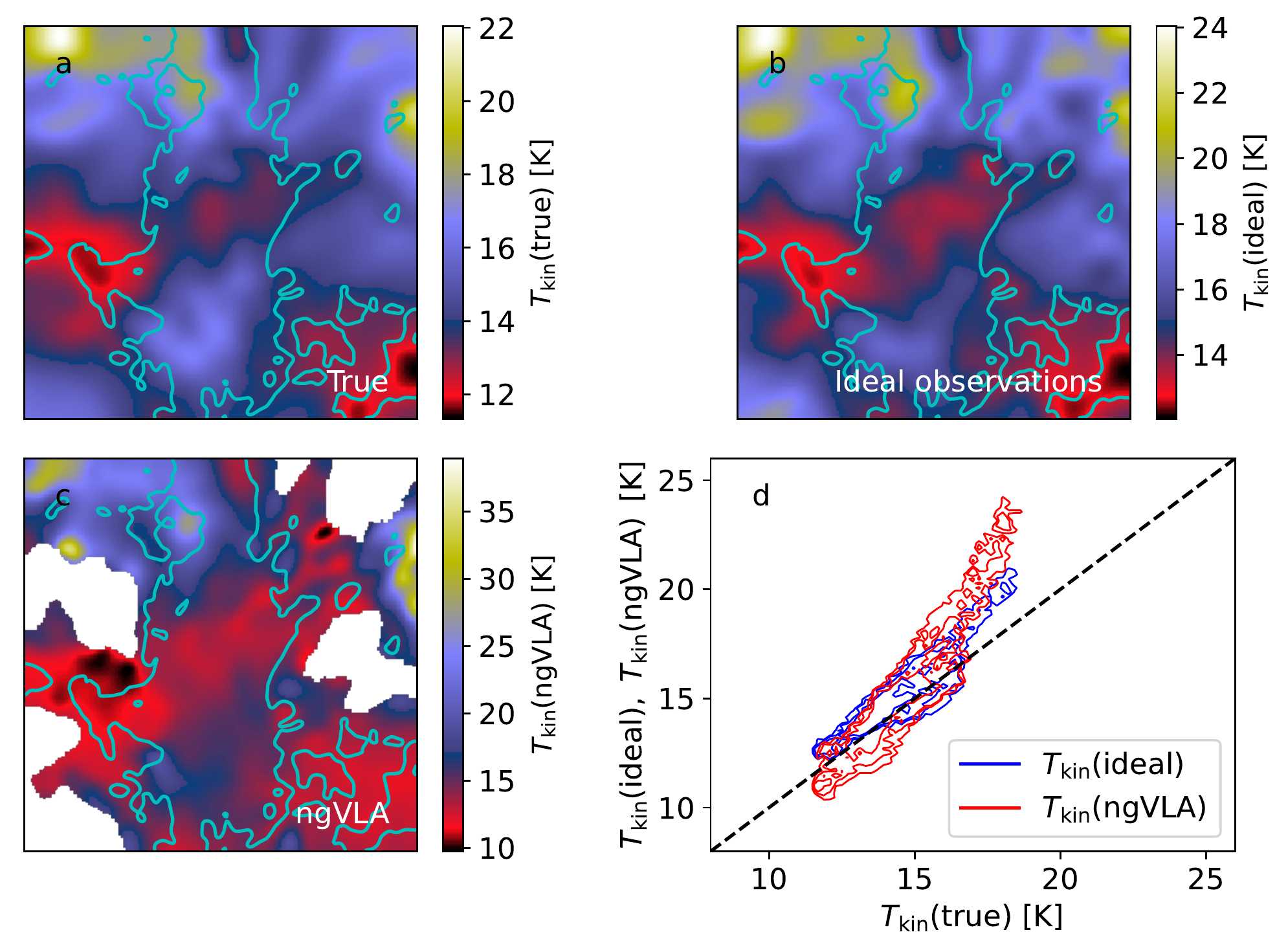}
\caption{
Comparison of kinetic-temperature estimates. The frames show (a) true density-weighted and LOS-averaged
kinetic temperature $T_{\rm kin}^{\rm true}$ in the 3D model, (b) estimates $T_{\rm kin}^{\rm ideal}$
from ideal NH$_3$ observations, (c) estimates $T_{\rm kin}^{\rm ngVLA}$ from synthetic ngVLA
observations, and (d) correlations of $T_{\rm kin}^{\rm ideal}$ (blue contours) and $T_{\rm kin}^{\rm
ngVLA}$ (red contours) against $T_{\rm kin}^{\rm true}$. In frame d, the plot includes pixels $N({\rm
H}_2)>5\times 10^{22}$\,cm$^{-2}$ (corresponding to the cyan contour in frames a-c), and the contours
are drawn at 20, 50, and 80 percent of the maximum value. All data are shown at 3$\arcsec$ resolution.
}
\label{fig:plot_Tkin}
\end{figure}

We used hyperfine fits of the ${\rm NH_3}$(1, 1) line and Gaussian fits of the ${\rm NH_3}$(2, 2) line,
assuming a single velocity component. Figure~\ref{fig:plot_Tkin} compares the true density-weighted
kinetic temperature $T_{\rm kin}^{\rm true}$ of the model cloud with the values derived from ideal and
from synthetic ngVLA ammonia observations. All data are convolved to 3$\arcsec$ resolution.

The ideal ${\rm NH_3}$ observations result in correct values to within a couple of degrees, with a
small positive bias across all temperatures. In the centre of the map, the presence of multiple
velocity components can also contribute to the differences. For synthetic ngVLA observations, the locus
coincides with that of ideal observations. However, there is some positive bias at higher temperatures
(mostly lower column densities). The maps suggest that these deviations are associated with the border
regions with lower S/N, especially for ${\rm NH_3}$(2,2), and large-scale effects from the filtering
of extended emission.

\begin{figure*}
\centering
\sidecaption
\includegraphics[width=12cm]{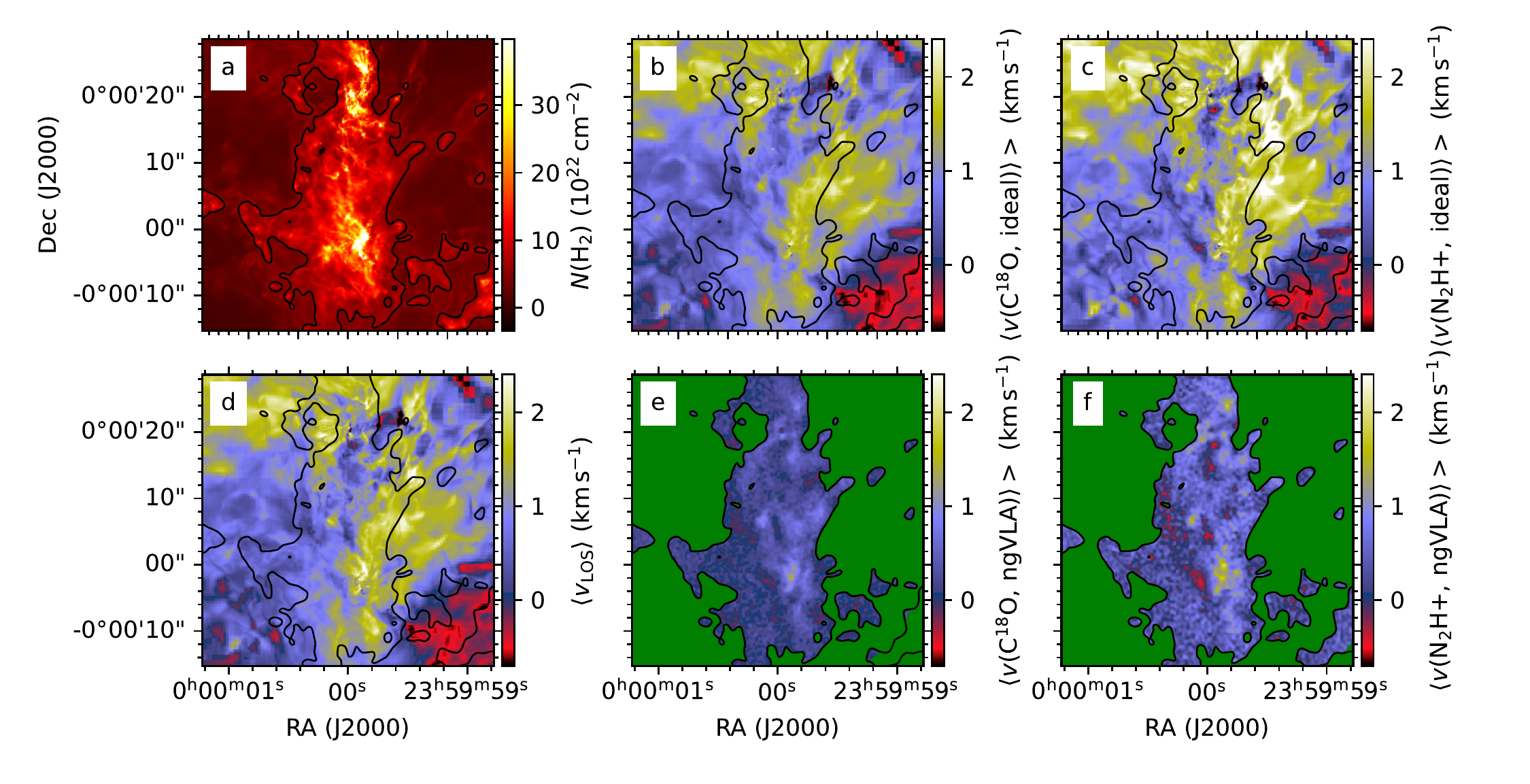}
\caption{
Comparison of mass-weighted average LOS velocity $\langle v_{\rm LOS} \rangle$ of the model cloud and
the mean radial velocity $\langle v \rangle$ from ideal observations or from synthetic ngVLA
observations. Frames show (a) true column density, (b) $\langle v \rangle$ in ideal ${\rm C^{18}O}$
observations, (c) $\langle v \rangle$ in ideal ${\rm N_2 H^{+}}$ observations, (d) $\langle v_{\rm LOS}
\rangle$ convolved to 0.55$\arcsec$ resolution, (e) $\langle v \rangle$ in synthetic ngVLA ${\rm
C^{18}O}$ observations, and (f) $\langle v \rangle$ in synthetic ngVLA ${\rm N_2 H^{+}}$ observations.
The contour is drawn at column density $N({\rm H}_2)=5\times 10^{22}$\,cm$^{-2}$, and regions of lower
column density are masked in frames e-f.
}
\label{fig:plot_VLOS_1}
\end{figure*}

\subsection{Cloud kinematics} \label{sect:kinematics}

Regarding the cloud kinematics, we look first at the differences between the true mass-weighted average
LOS velocity in the model cloud and the mean velocities estimated from line observations. Second, we
look at infall indicators and how they are related to the actual infall motions in the model cloud.

\subsubsection{Small-scale velocity field} \label{sect:LOS_velocity}

We compare the observed mean line velocities $\langle v \rangle$ to the mass- and abundance-weighted
LOS velocity $\langle v_{\rm LOS} \rangle$. The observed $\langle v \rangle$ are calculated as the
intensity-weighted average radial velocity over the line profiles, both for the ideal spectra and the
synthetic ngVLA spectra. The parameter $\langle v_{\rm LOS} \rangle$ is read directly from the 3D model
cloud. We include in its definition also the abundance variations, so that ideally $\langle v_{\rm LOS}
\rangle \approx \langle v \rangle$ for optically thin emission.

Figure~\ref{fig:plot_VLOS_1} compares the observed $\langle v \rangle$ of the ${\rm C^{18}O}$ and ${\rm N_2
H^{+}}$ spectra to the actual $\langle v_{\rm LOS} \rangle$ values in the model. The observed $\langle v
\rangle$ values differ from $\langle v_{\rm LOS} \rangle$ by up to $\sim$1\,km\,s$^{-1}$. These can be
caused by temperature variations that affect the observed spectra but not the $\langle v_{\rm LOS}
\rangle$ values. A second explanation are radiative transfer effects, because both ${\rm C^{18}O}$ and
${\rm N_2 H^{+}}$ reach optical depths slightly above one.
The radial velocities derived from the two species are clearly correlated, but also show differences
below 1\,km\,s$^{-1}$.  Although synthetic ngVLA observations may be affected by the interferometric
filtering, their $\langle v \rangle$ values are still clearly closer to the velocities of the ideal
spectra than to the mass-weighted average velocities $\langle v_{\rm LOS} \rangle$ of the model cloud.

\begin{figure*}
\centering
\sidecaption
\includegraphics[width=12cm]{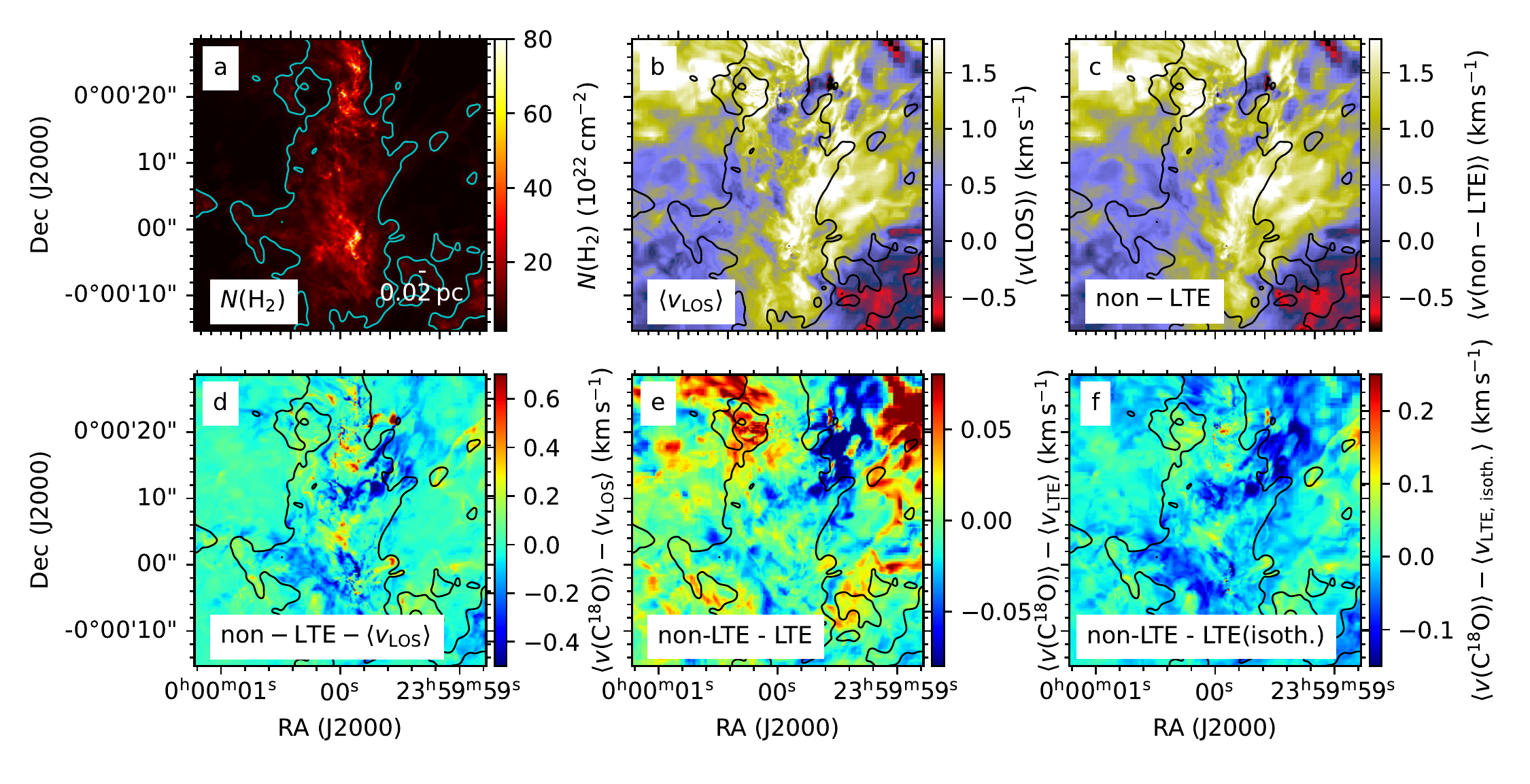}
\caption{
Comparison of radial-velocity estimates from ideal C$^{18}$O line observations. The frames show: (a)
true column density of the model, (b) true mass-weighted average LOS velocity $\langle v_{\rm LOS}
\rangle$ of the model, (c) mean radial velocity in non-LTE spectra, (d) difference between the radial
velocity of the non-LTE spectra and $\langle v_{\rm LOS} \rangle$, (e) velocity difference between
non-LTE and LTE spectra (including $T_{\rm kin}$ variations), and (f) velocity difference between
non-LTE and LTE spectra, the latter assuming a constant kinetic temperature of $T_{\rm kin}=15$\,K. All
data are used at full resolution, without beam convolution. 
}
\label{fig:plot_VLOS_2}
\end{figure*}

Figure~\ref{fig:plot_VLOS_2} compares radial velocities of different types of idealised spectra. It
shows that the differences between non-LTE spectra and $\langle v_{\rm LOS} \rangle$ are explained more
by the variations in kinetic temperature (frame f) than by non-LTE effects (frame e). Together these
account for about for half of the differences to $\langle v_{\rm LOS} \rangle$, self-absorption and other
radiative-transfer effects providing further contributions.

\begin{figure*}
\centering
\sidecaption
\includegraphics[width=12cm]{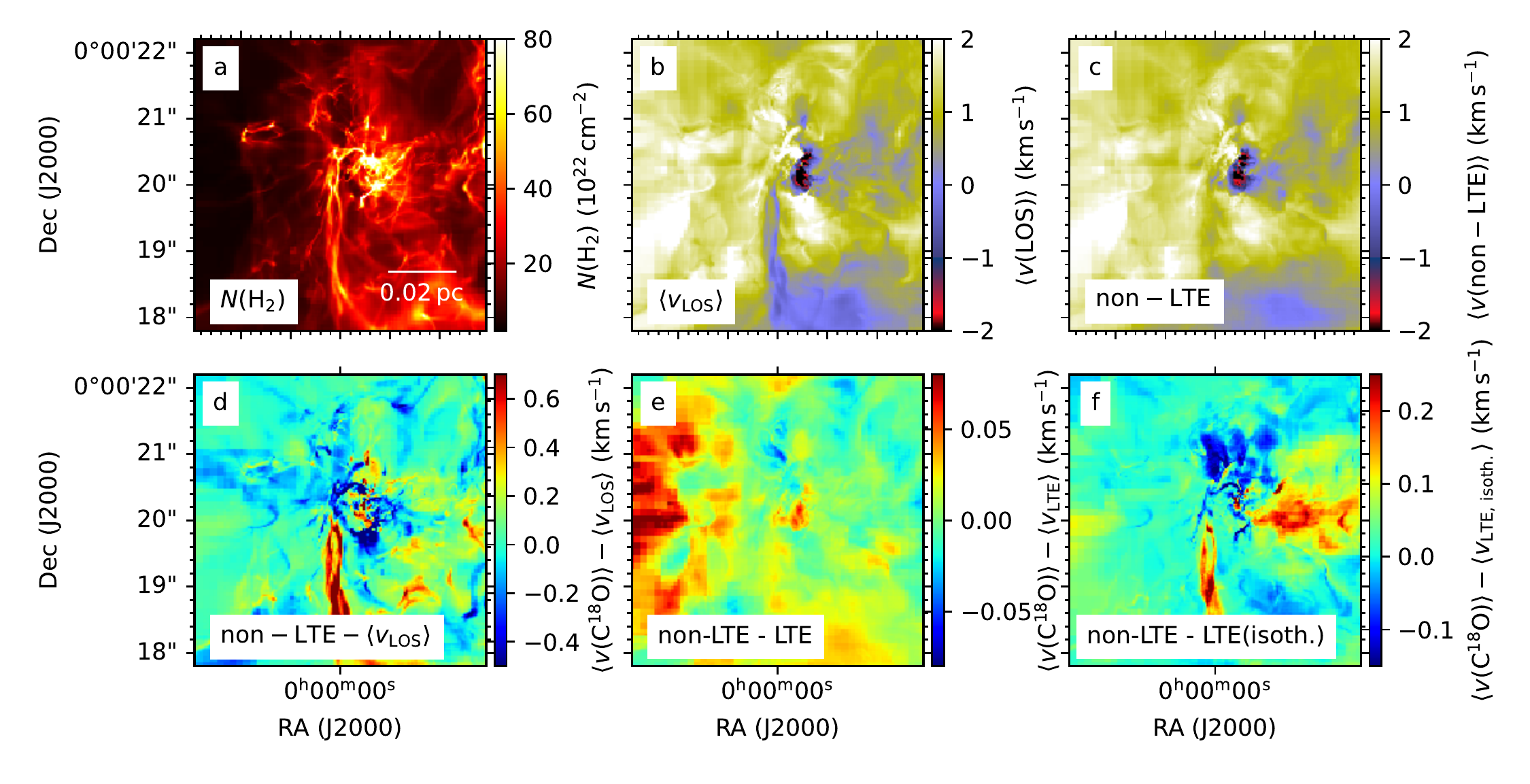}
\caption{
As Fig.~\ref{fig:plot_VLOS_2} but zooming to the northern clump. 
}
\label{fig:plot_VLOS_2_ZOOM}
\end{figure*}

Figure~\ref{fig:plot_VLOS_2_ZOOM} shows the same data in the northern core. There are several
interesting kinematic features, such as a linear vertical filament (south of the central core, with
length of $\sim$0.05\,pc and a width of ~0.001\,pc) and spiralling structures close to the core centre,
which itself shows velocity differences of more than 4\,km\,s$^{-1}$ between its northern and southern
parts. The area of the vertical filament shows a particularly large difference between $\langle v_{\rm
LOS} \rangle$ and the $\langle v \rangle$ values derived from the non-LTE spectra. The structure almost
completely disappears in the latter, because of its low excitation compared to other material along the
LOS.

\begin{figure}
\centering
\includegraphics[width=9cm]{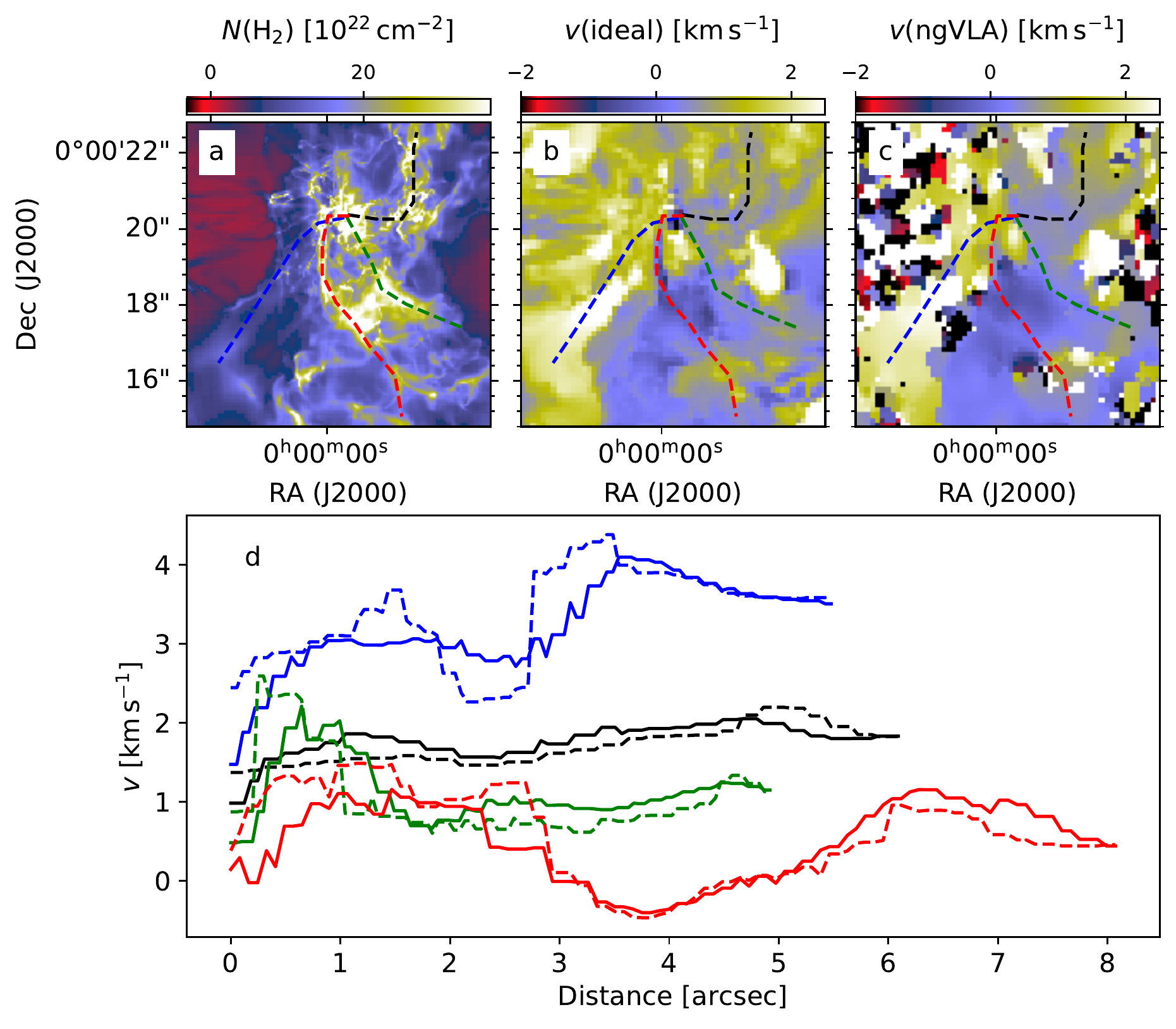}
\caption{
Comparison of radial-velocity estimates from ideal and synthetic ngVLA observations of ${\rm
H^{13}CO^{+}}$(1-0). Frames a-c show, respectively, the true column density, the radial velocity from
ideal observations, and the radial velocity from synthetic ngVLA observations. Frame d shows the radial
velocity along the four paths (starting at the core location) indicated in frames a-c. The radial
velocity from ideal observations is plotted with solid lines and the velocity from synthetic ngVLA
observations with dashed lines.
}
\label{fig:H13COp_velocity}
\end{figure}

Figure~\ref{fig:H13COp_velocity} compares further the ideal and synthetic ${\rm H^{13}CO^{+}}$
observations of the northern core. The radial velocities are taken from Gaussian fits with a single
velocity component\footnote{Compared to direct intensity-weighted mean velocities, the single-component
Gaussian fits are here better in extracting the velocity of the dominant LOS emission structure.}, and
they are traced along four paths starting at the core location. The ngVLA observations follow the
results of ideal observations, typically with a precision to a fraction of 1\,km\,s$^{-1}$. Towards the
core, limited by the resolution of the observations, the velocity gradients increase up to
$\sim$50\,km\,s$^{-1}$\,pc$^{-1}$. However, large gradients are seen also further out, as the selected
path crosses different LOS emission regions. The complexity of the LOS structure is demonstrated
further in Appendix~\ref{app:LOS_emission}, where show plots of the 3D density cube and the
position-position-velocity (PPV) cube derived from synthetic ngVLA observations of the $^{13}$CO line.

\subsubsection{Large-scale velocity field} \label{sect:PCA}

Principal component analysis (PCA) uses the eigenvalue decomposition of the spectral observations to
examine velocity fluctuations as a function of scale \citep{Heyer_1997} This usually results in a
scaling relation $\delta v \sim R^{\alpha}$, which can be further related to the energy power spectrum
\citep{Brunt_Heyer_2002, Brunt_Heyer_2013}.

\begin{figure}
\centering
\includegraphics[width=9cm]{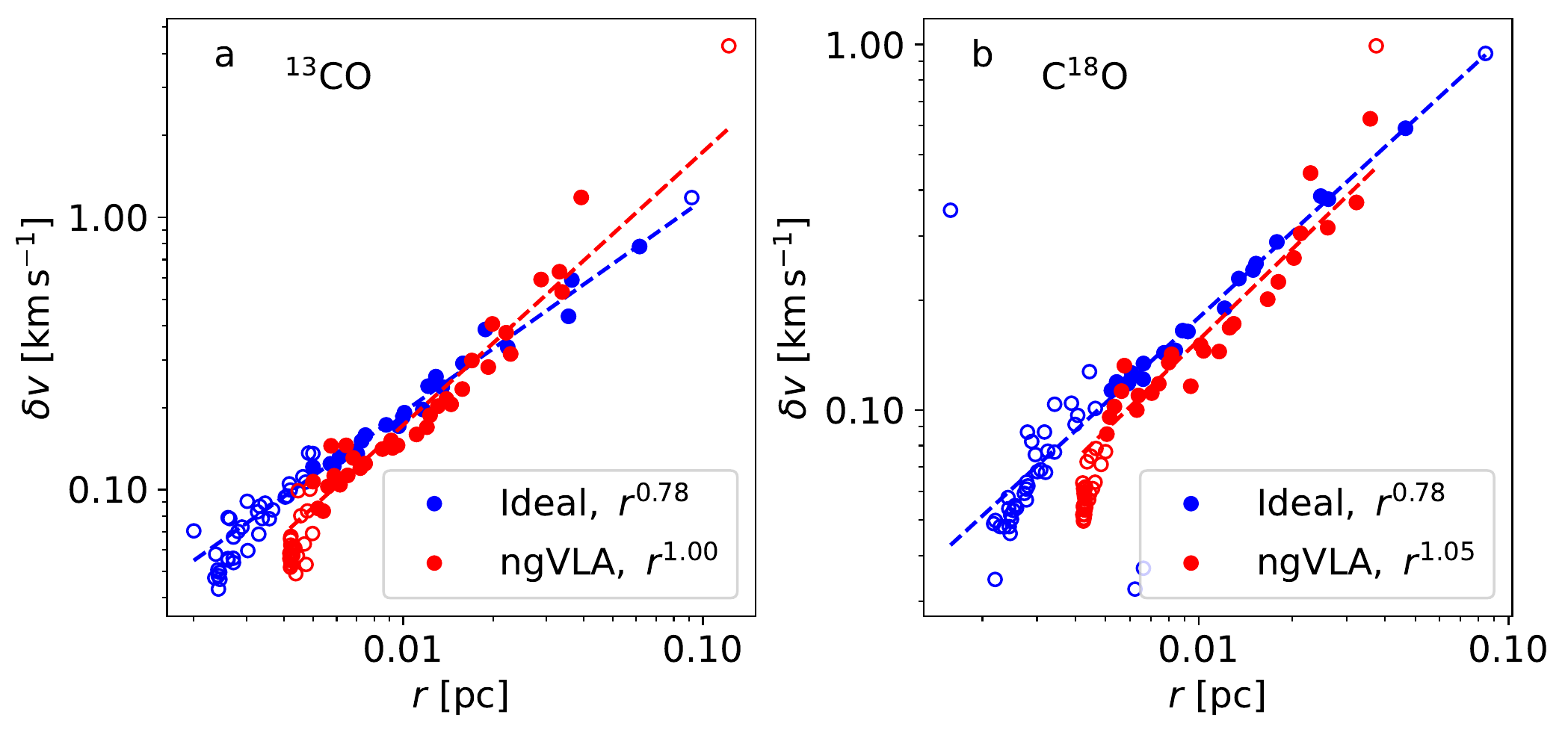}
\caption{
Results of PCA analysis of ideal and synthetic ngVLA observations of the $^{13}$CO and ${\rm C^{18}O}$
lines. The quoted exponents of the scaling relation correspond to least-squares fits to the filled
symbols.
}
\label{fig:plot_PCA}
\end{figure}

Figure~\ref{fig:plot_PCA} shows the calculated scaling relations for $^{13}$CO and C$^{18}$O, for an
area of $27\arcsec\times 41\arcsec$ centred on the filament. In ideal observations, both molecules give
the same value of 0.78 for the exponent of the scaling relation, in spite of the optical depths
differing by a factor of several (fractional-abundance ratio of ten). The slopes are significantly
steeper for the synthetic ngVLA observations, with little difference between the $^{13}$CO and ${\rm
C^{18}O}$ lines. The downturn and truncation of the relationships around 0.004\,pc is due to the beam
size ($0.55\arcsec \sim 0.001$\,pc).

\subsubsection{Infall indicators} \label{sect:infall}

The main structure of the model cloud consists of the northern and the southern clumps. These have
complex 3D velocity fields, but the net flow of gas is directed towards the clump centres and
many individual cores. 
However, observations provide information only of the LOS velocities. To characterise the corresponding
LOS motions in the 3D model, we define an inflow index $\xi$ with the equation
\begin{equation}
\xi =  \langle  {\rm sign}(x_0-x) \, \, n^{\prime}(x) \,\, (v(x)-v_0)  \rangle
\, \,  /  \, \, \langle n^{\prime}(x) \rangle,
\label{eq:infall_xi}
\end{equation}
where $v(x)$ and $n(x)$ are the velocity and the density along the LOS coordinate $x$. The location of
the main LOS density maximum is $x_0$, the radial velocity at that position is $v_0$, and the sign is
selected so that infall towards the maximum corresponds to positive $\xi$ values, the $\xi$ parameter
having units of velocity. The averaging can be done over the whole LOS but, to measure gas motions above
a given density threshold $n_0$, we use $n^{\prime}=\max(0, n(x)-n_0)$.

\begin{figure}
\centering
\includegraphics[width=8.8cm]{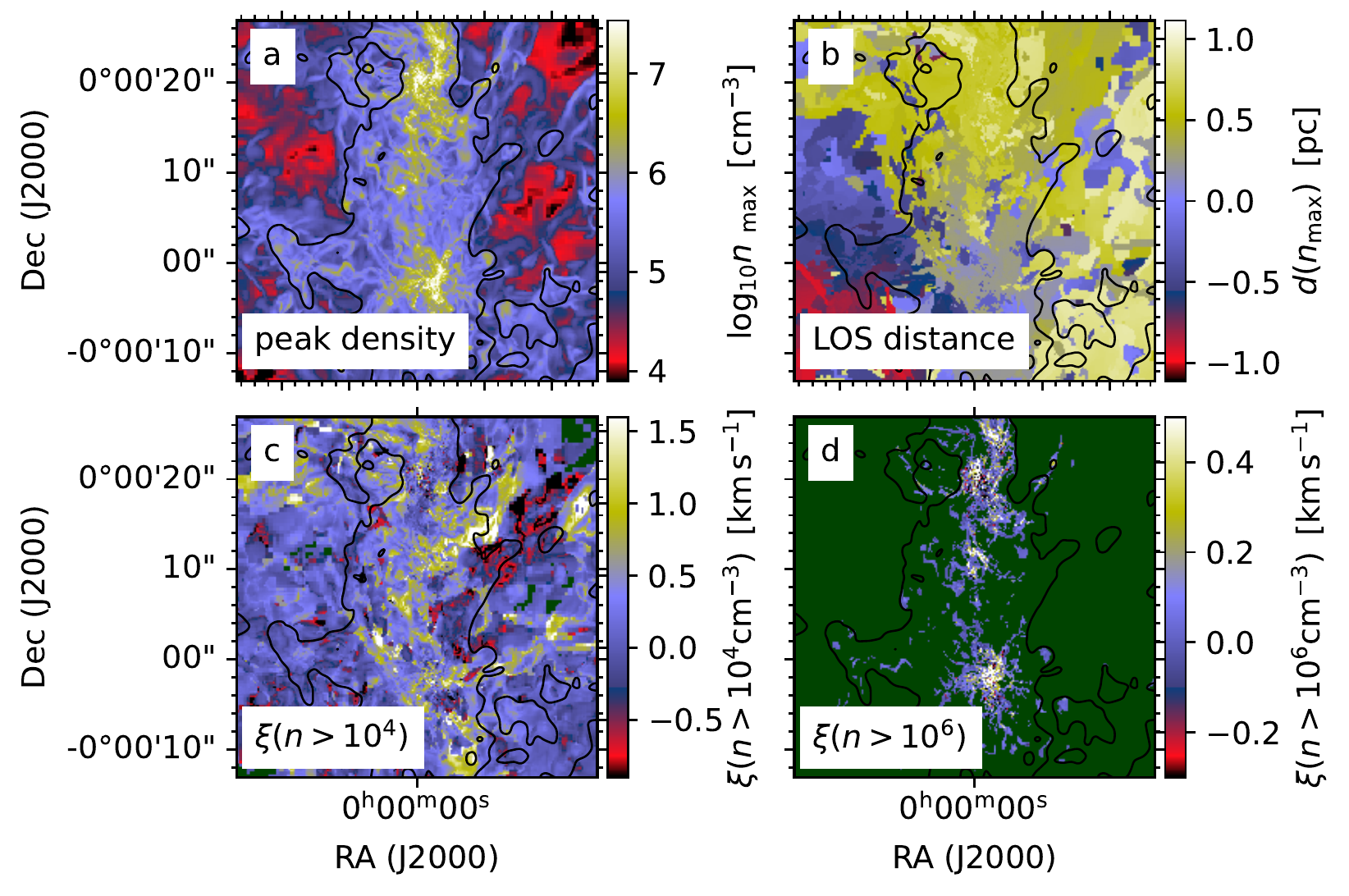}
\caption{
Maps of infall index $\xi$ computed from the 3D model. Upper frames show the peak density along each
LOS (frame a) and the distance (relative to the cloud centre) to that location along the LOS (frame b).
The lower frames show the $\xi$ maps for density thresholds $n_0=10^4$\,cm$^{-3}$ and
$n_0=10^6$\,cm$^{-3}$. The dark green colour corresponds to masked areas where the maximum density
along the LOS falls below the density threshold $n_0$.
}
\label{fig:infall_index}
\end{figure}

\begin{figure}
\centering
\includegraphics[width=8.8cm]{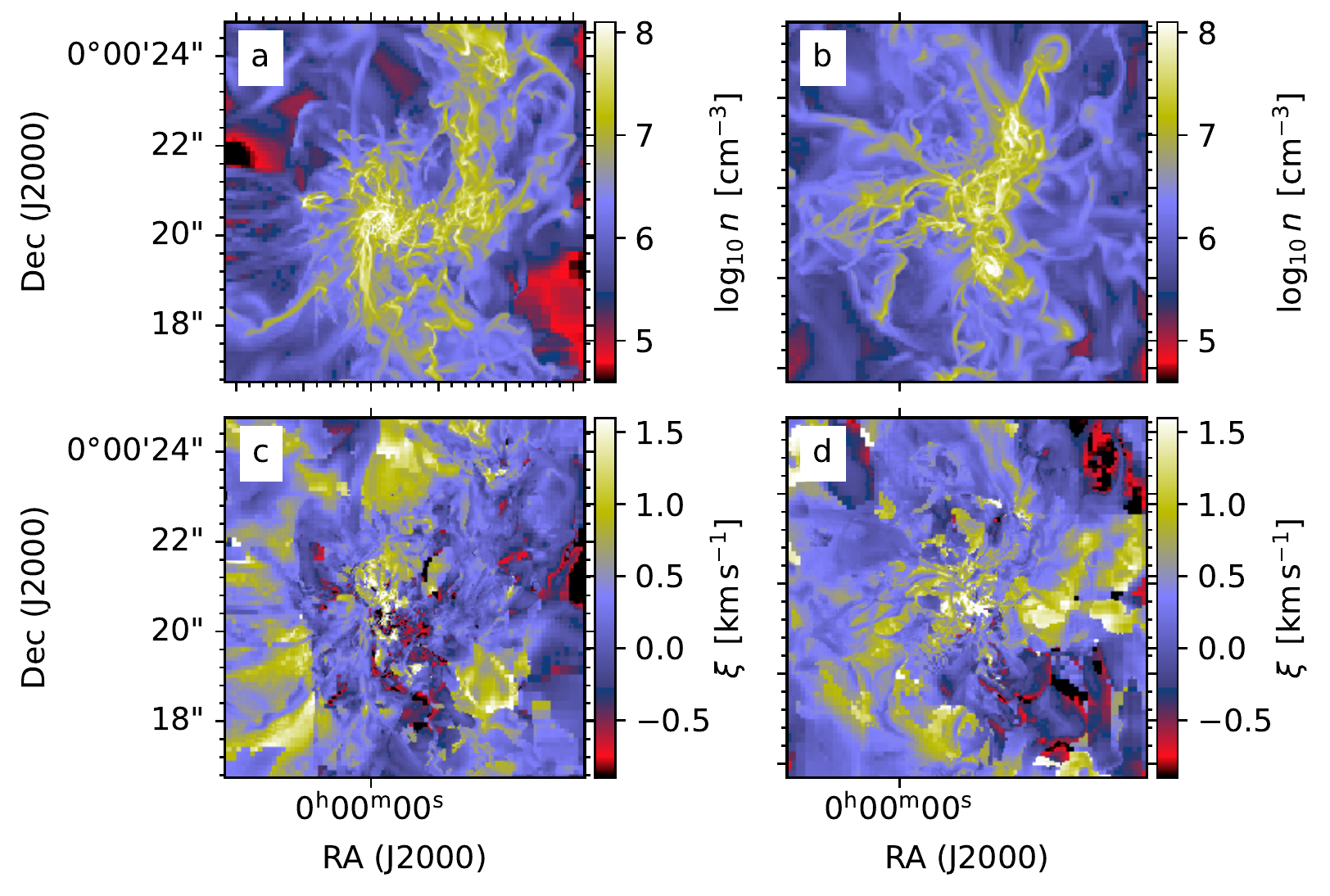}
\caption{
Upper frames show maps of the peak LOS density towards the northern (frame a) and the southern (frame
b) core. The lower frames show the corresponding maps of the infall index $\xi$, which are computed
using a threshold value of $n_0=10^4$\,cm$^{-3}$.
}
\label{fig:infall_index_zoom}
\end{figure}

Figure~\ref{fig:infall_index} shows the $\xi$ values for the whole filament area, for the threshold
values of $n_0=10^4$\,cm$^{-3}$ and $n_0=10^6$\,cm$^{-3}$. The high-density gas shows almost
exclusively positive $\xi$ values (especially in frame d), indicating the systematic contraction of the
structure. 
When the clumps are examined in more detail, larger $\xi$ values can be resolved towards the central
cores at sub-arcsecond scales (Fig.~\ref{fig:infall_index_zoom}). The mean values of $\xi$ are clearly
positive, but there are also small intertwined areas with negative $\xi$ values. When observed at lower
resolution, these can be thus expected to dilute the overall infall signature.

In line observations, infall regions might be identified with analysis of the line profiles, with the
comparison of optically thin and thick tracers. This presumes that the excitation temperature
increases towards the infall centre, which then leads to asymmetric self-absorption and the net
emission of optically thick lines moving to lower radial velocities and results in a blueshifted
profile. The assumption is likely to be correct for isolated, symmetric, and centrally concentrated
structures, rendering it very useful in the study of isolated cores. However, if the line profile
contains emission from completely separate LOS structures (with different radial velocities and
excitation conditions), the line profile cannot be expected to be closely correlated with the infall
motions of individual cores

We examined this question using a parameter
\begin{equation}
\delta v = \frac{v_{\rm thin} - v_{\rm thick}}{\Delta v_{\rm thin}}
\label{eq:infall_dv}
\end{equation}
that compares the radial velocities of one optically thin and one optically thick line
\citep[cf.][]{Mardones1997}, adopting a convention where {\em positive} values of $\delta v$ indicate
infall or collapse. We calculated $\delta V$ for three pairs of lines: ${\rm C^{18}O}$(1-0) and
$^{13}$CO(1-0), ${\rm H^{13}CO^{+}}$(1-0) and HCO$^{+}$(1-0), ${\rm N_2 H^{+}}$(1-0) and
HCO$^{+}$(1-0). Two alternative sets of input values were used in Eq.~(\ref{eq:infall_dv}). In the
first case, the velocities and the line width were obtained from Gaussian fits with a single velocity
component. In the second case, we examined directly the channels with optically thin emission above a
given brightness-temperature threshold and estimated $v_{\rm thin}$ and $v_{\rm thick}$ directly as
weighted averaged over these channels.

\begin{figure*}
\centering
\sidecaption
\includegraphics[width=12cm]{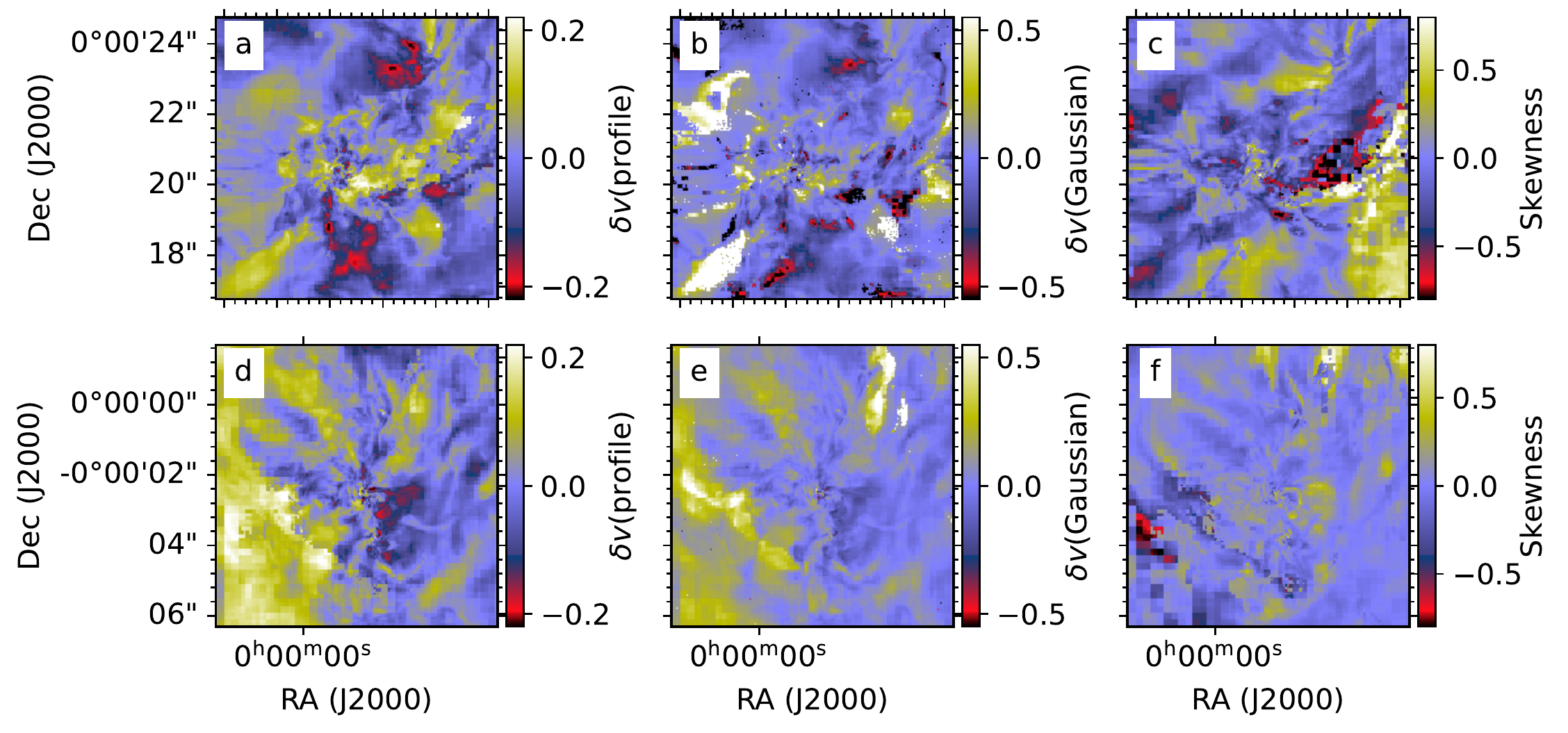}
\caption{
Collapse indicators $\delta V$ based on ideal ${\rm C^{18}O}$ and ${\rm ^{13}CO}$ observations. The
upper and lower frames show, respectively, the northern and the southern clumps, for the same area as
in Fig.~\ref{fig:infall_index_zoom}. Frames a and d show $\delta v \rm (profile)$ that is computed from
channels around the peak of the ${\rm C^{18}O}$ spectrum. Frames b and e show the corresponding $\delta
v(\rm Gaussian)$ that is based on line parameters from Gaussian fits. For comparison, frames c and f
show the skewness of the ${\rm ^{13}CO}$ line profiles.
}
\label{fig:LOS_collapse_LOC}
\end{figure*}

Figure~\ref{fig:LOS_collapse_LOC} shows the results for ideal ${\rm C^{18}O}$(1-0) and $^{13}$CO(1-0)
observations. 
For comparison, we plot in Figure~\ref{fig:LOS_collapse_LOC} also the skewness of the $^{13}$CO(1-0)
profile (for channels around the peak of the ${\rm C^{18}O}$(1-0) spectrum).

The values of $\delta v$ remain well below $\delta v \sim 2$ that has been considered significant in
studies of protostellar cores \citep{Mardones1997}. This is due to the very large total linewidths. The
two methods to compute $\delta v$ do mostly agree, but there are also areas where the Gaussian fits
lead to clearly different or even opposite results. This is not surprising, considering the complexity
of some of the spectral profiles. There is very little correlation between the $\delta v$ and the $\xi$
values of Fig.~\ref{fig:infall_index_zoom}, and there are marked differences also between the $\delta
v$ maps derived from ideal observations (Fig.~\ref{fig:LOS_collapse_LOC}) or from the synthetic ngVLA
observations (not shown).

The scenario where $\delta v$ is able to probe collapse (an isolated core with centrally peaked $T_{\rm
ex}$ distribution and a symmetric velocity field) will also result in positive kurtosis in the profiles
of optically thick lines. The correlation between kurtosis and $\delta v$ is weak for the ideal
observations but becomes significant in the synthetic ngVLA observations. This shows that these
statistics are not efficient in localising regions of infall motion, not only because of the more
complex physical situation but also because they are sensitive to systematic errors that affect the
data at larger scales (e.g., self-cancellation due to the extended emission from optically thick
species).

We examine the LOS emission further in Appendix~\ref{app:LOS_emission}. There the spectra are seen
consist of emission from a number of disjoint density peaks, making $\delta v$ insensitive to actual
infall motions in any single LOS density peak. We return to this question also in the following
discussion (Sect.~\ref{disc:infall}).

\section{Discussion} \label{sect:discussion}

We have investigated the use of line observations for studies of star-forming clouds, comparing ideal
and synthetic observations to the known properties of the model cloud. In particular, we have simulated
observations for the planned ngVLA radio interferometer. We discuss below the main result.

\subsection{Cloud model and synthetic observations}

The model cloud has a mass of some 3000\,M$_\sun$ in a volume of (4\,pc)$^3$. It has thus a mean
density of nearly $n({\rm H}_2)=700$\,cm$^{-3}$ and is capable of star formation up to high-mass stars
\citep{Haugbolle2018}. We concentrated on the densest part of the, where the column densities reach up
to $N({\rm H_2}) \sim 10^{24}$\,cm$^{-2}$. The target is thus similar to a small filamentary infrared
IRDC \citep[e.g.][]{JimenezSerra2014, Barnes2021, Liu2022}. As shown by Fig.~\ref{fig:colden},
the appearance of a single filament can easily result from projection effects in a turbulent cloud
\citep{Juvela2012b}. It is noteworthy that in two out of the three orthogonal view direction the same
disjoint 3D regions align to form a single filamentary structure. They also have a velocity difference
only of a couple of km\,s$^{-1}$ and thus cannot be easily separated even in velocity space. The cloud
also includes several hundred newly-born stars that provide local heating and increase the LOS
temperature variations. The average temperatures are low (with a mode of 14.3\,K) but exceed 100\,K in
small regions. Therefore, the model provides a good analogue for a high-mass star forming region with
some of its observational challenges.

Nevertheless, there are limitations to the realism of the cloud model. First, it does not include a
detailed description of molecular outflows that could considerably further complicate the kinematics.
Second, the chemical abundances were set using an ad-hoc density dependence, instead of detailed
time-dependent chemical modelling. Third, the computed dust temperatures were used as a proxy for the
gas temperatures, instead of direct computation of the gas thermal balance. This is justified at high
densities (Sect.~\ref{sect:RT}) but is increasingly inaccurate when densities fall below $\sim
10^5$\,cm$^{-3}$. 

The gas heating varies depending on the local conditions, also in ways that are not considered in our
model. Cosmic-ray fluxes in excess of $\xi=10^{-14}$\,s$^{-1}$ have been observed towards some
high-mass star-forming regions. These are much higher than the typical values in dense clouds ($\sim
10^{-17}$\,s$^{-1}$ and above, \citep{Caselli1998,Gerin2010}). However, the high rates are found
preferentially in regions of low density and low molecular abundances \citep{Bayet2011}. Cosmic rays
are attenuated only by thick columns of gas ($N({\rm H_2})\sim 10^{23}$\,cm$^{-2}$) but are also
affected by the magnetic fields. \citet{Owen2021} investigated cosmic-ray propagation in filamentary
clouds and the resulting effects on gas temperatures. The effect reaches almost 20\,K, when the
cosmic-ray energy density is increased a factor of ten above its normal Milky-Way value. However, those
calculations correspond to lower densities and therefore do not include the gas-dust coupling. In
models, where the gas would be heated to similarly high temperatures via photoelectric effect, the
collisions are able to bring the gas temperature down to within some degrees of the dust temperature
\citep{Goldsmith2001, Juvela2011}. Therefore, although the gas temperatures in our model are not
calculated precisely, they are still representative of the temperature fields in real clouds.

More locally, the formed high-mass stars create photon dominated regions (PDRs) and the young stars can
produce X-rays with distinct effects on the chemistry \citep{Hollenbach1997, Spaans2005,
Meijerink2006}. The modelling of the detailed temperature structure and chemistry of PDRs is clearly
beyond the scope of this paper. The MHD simulation contains a few bona fide massive stars, but, in
their surroundings, the predicted line emission will be more reminiscent of earlier evolutionary
stages. The selected angular resolution (1$\arcsec$ corresponding to 4000\,au) means that the synthetic
observations are not sensitive to structures at the scale of protostellar disks.

The above factors do not directly affect our main goal, the comparison of a 3D model and synthetic
observations made of that model. Both the column-density and temperature distributions are similar to
those found in real IRDCs. Furthermore, the synthetic line observations were calculated with full
radiative transfer calculations, which take into not only the varying densities and kinetic
temperatures but also deviations from LTE conditions.

\subsection{Column densities}

As long as the observed lines are not completely optically thick, column densities can be estimated
using a single line with an assumed excitation temperature or with two or more lines with an assumed or
measured opacity ratio. The cases of the ${\rm N_2 H^{+}}$ and ${\rm NH_3}$ column densities fall into
the second category, making use of the known the optical-depth ratios of the hyperfine components. All
our column-density estimates include the assumption of a homogeneous medium, and the resulting errors
can be expected to increase as the optical depths and the density and kinetic-temperature variations
increase.

Because of the low average kinetic temperatures, the brightness temperatures are in our simulations
typically only of the order of 10\,K. High sensitivity is therefore required to reliably measure the
line intensities outside the densest cores, especially in the case of the ${\rm N_2 H^{+}}$ and ${\rm
NH_3}$ satellite lines. Of these two species, the ${\rm NH_3}$ integrated intensities are 2-3 times
larger (Fig.~\ref{fig:W}). The largest differences between ${\rm N_2 H^{+}}$ and ${\rm NH_3}$ are
caused, however, from the almost factor of four difference in frequencies. As a result, S/N of ${\rm
NH_3}$ is much higher but the angular resolution is correspondingly lower (2.43$\arcsec$ vs.
0.64$\arcsec$; Fig.~\ref{fig:colden_maps}). Both species provide accurate column density estimates in
ideal observation. In the synthetic ngVLA observations, ${\rm N_2 H^{+}}$ shows a larger dispersion
(for a significantly higher angular resolution) and systematically lower values
(Fig.~\ref{fig:plot_colden_cor}). The low values are naturally explained by the filtering of extended
emission, which is evident for example in Fig.~\ref{fig:W}. 
Similar to ${\rm N_2 H^{+}}$, reliable observations of ${\rm C^{18}O}$ and ${\rm H^{13}CO^{+}}$ are
limited to regions of highest column density ($N({\rm H_2})\ga5\cdot 10^{22}$\,cm$^{-2}$).  Previous
observations of IRDCs have shown that e.g. H$^{13}$CO$^{\rm +}$(1-0) emission matches continuum
emission quite well in dense regions \citep{Liu2022}.

For the ngVLA Core antenna configuration, the maximum recoverable scales are almost 80$\arcsec$ at the
frequency of the ${\rm NH_3}$(1,1) line and slightly less than 20$\arcsec$ at the frequency of the
${\rm N_2 H^{+}}$(1-0) line. Our cloud falls between these scales, the filament ($N({\rm H}_2)>5\times
10^{22}$\,cm$^{-2}$) extending over an area of some $40 \times 10$ arcsec. As pointed out in ngVLA
memos\footnote{https://library.nrao.edu/ngvla.shtml, memo \#14}, at scales above 5$\arcsec$ it becomes
important for image fidelity and sensitivity to combine interferometric observations with data from a
large single-dish telescope. We return to the importance of single-dish data in 
Sect.~\ref{sect:singleDish}.

We estimated column densities also with ${\rm C^{18}O}$(1-0) spectra and the assumption of a constant
excitation temperature. For ideal observations, this resulted in fairly accurate estimates, the
break-up of the assumption of optically thin lines causing significant underestimation only at the
highest column densities (Fig.~\ref{fig:plot_colden_cor}a). However, in our cloud model also the
kinetic temperature tends to decrease with increasing density, which could explain part of the bias.
In the case of synthetic ngVLA observations, the column densities are underestimated, similar to the
${\rm N_2 H^{+}}$ results, confirming that the bias is common for observations towards the upper end of
the ngVLA frequency range and related to the lack of information of the lowest spatial frequencies.

Column densities were calculated using Gaussian fits or, in the case of hyperfine structure, a set of
Gaussians for the hyperfine components but only a single velocity component. It is difficult to
predict, how much error the Gaussian approximation produces, because this can depend even on the
initial values used in the fits. The fitted component might cover multiple velocity components or could
converge to just one of those. The effect on ${\rm C^{18}O}$ column-density estimates is easy to
understand, because these are directly proportional to the integrated line intensity. Even in the case
of multi-modal spectra (examples shown in Fig.~\ref{fig:fit_c18o_CASA}), the differences between the
one-component and two-component fits were small at high column densities
(Fig.~\ref{fig:plot_colden_cor}). At lower column densities, the two-component fits resulted in larger
column densities, by up to a factor of two. Such differences could be expected in the centre of the
field, where the spectra show two equally strong velocity components.

The hyperfine structure of the ${\rm N_2 H^{+}}$ and ${\rm NH_3}$ lines was fitted using only a single
velocity component. Unlike in the simple Gaussian fits, these fits provide directly estimates for the
excitation temperature and optical depth. In this case, multi-component fits could easily lead to
unphysical solutions. For example, one components could have $T_{\rm ex}<T_{\rm bg}$, providing an
``absorption'' feature to help the fitting of some non-gaussian line profiles, or one could even have
$T_{\rm ex} \approx T_{\rm bg}$ combined with an arbitrary column density. Some unphysical solutions
could be automatically rejected or avoided by using suitable regularisation. However, fitting of
multi-component models to hyperfine spectra remains susceptible to degeneracies, and the automatic
selection of the physically most likely solution remains a challenge. In the purely technical sense,
the fits are still easy. For example, the maps of ideal observations in Fig.~\ref{fig:colden_maps}
contained some 5 million individual spectra but could fitted in just a few minutes.

\subsection{Estimates of kinetic temperature}

Figure~\ref{fig:plot_Tkin} showed kinetic-temperature estimates derived from the ${\rm NH_3}$(1,1) and
${\rm NH_3}$(2,2) spectra. The accuracy of the estimates obtained with synthetic ngVLA observations is
mostly comparable to that of ideal observations, with statistical noise of the order of 1\,K. However,
there is an increasing positive bias towards higher temperatures, which mirrors the behaviour in
Fig.~\ref{fig:plot_colden_cor}c for $N({\rm H}_2)$ at low column densities. This can be attributed
partly to imperfections of the fits (with a single velocity component) and the different optical depths
of the transitions, when combined with the large LOS temperature and density variations.
\citet{Shirley2015} estimated that the effective critical density of the ${\rm NH_3}$(2,2) is at
10-20\,K temperatures about two times higher than that of ${\rm NH_3}$(1,1). In the $T_{\rm kin}$
analysis, the ${\rm NH_3}$(2,2) spectra were fitted using fixed radial velocities obtained from the
${\rm NH_3}$(1,1) hyperfine fits. If the radial velocities of the ${\rm NH_3}$(2,2) line are kept as
free parameters, the appearance of Fig.~\ref{fig:plot_Tkin} changes only little, which shows that the
two transitions are still essentially probing the same volume of gas. The temperature determination may
also be influenced by imperfections in the interferometric observation and the data reduction, larger
self-cancellation in ${\rm NH_3}$(1,1) observations naturally leading to positive bias in $T_{\rm
kin}$ estimates. The S/N of the synthetic ngVLA observations was clearly sufficient to
map kinetic temperature over large cloud areas, beyond the $N({\rm H}_2)=5\times 10^{22}$\,cm$^{-2}$
threshold used in Fig.~\ref{fig:plot_Tkin}.

\subsection{Velocity fields}  \label{sect:velocity_fields}

Analysis in Sect.~\ref{sect:LOS_velocity} showed that the mean radial velocity of observed spectral
lines can significantly differ from the actual mass-weighted mean radial velocity in the model cloud.
For ideal observations, which were computed at the full resolution of the model, the differences were up
to $\pm$1\,km\,s$^{-1}$. This corresponds to a large fraction of the total range of velocities.
The total velocity dispersion of our model cloud is $\sim$4\,km\,s$^{-1}$, which is also consistent
with observations of many IRDCs \citep{Li2022, Liu2022}.
The large differences between the observed and true (mass-weighted) velocities are limited to small
regions, typically with large volume densities and large optical depths. The synthetic ngVLA
observations and ideal line observations show differences at two distinct scales. At the smallest
scales, close to the resolution of the observations, these result from observational noise. In the area
with $N({\rm H_2})>5\times 10^{22}$\,cm$^{-2}$, the rms differences are quite small,
$\sim$0.1\,km\,s$^{-1}$ and $\sim$0.15\,km\,s$^{-1}$ for ${\rm C^{18}O}$ and ${\rm N_2 H^{+}}$,
respectively. These values were measured at scales below 4$\arcsec$, by filtering out the larger-scale
signal in the velocity maps. The absolute differences could be several times higher, but these are
limited to larger spatial scales ($>4\arcsec$) and are observed to be similar for both molecules.
Therefore, based on the similar frequencies of the lines, these are most likely related to the
filtering of large-scale emission and other imperfections in the imaging procedures.

Further investigations showed that the differences between the radial velocity of spectra and the true
gas velocities are caused by temperature variations and optical-depth effects, which change the
relative contribution of different LOS regions. The non-LTE effects (i.e., $T_{\rm ex}$ variations in
addition to $T_{\rm kin}$ variations) had a smaller impact
(Figs.~\ref{fig:plot_VLOS_2}-\ref{fig:plot_VLOS_2_ZOOM}).  
The synthetic ngVLA observations traced well the large-scale velocity structures, from the beam size up
to at least $\sim 4\arcsec$. Thanks to the hyperfine structure, the velocity determination is
particularly accurate with ${\rm N_2 H^{+}}$ observations (Fig.~\ref{fig:plot_VLOS_1}).

Zooming into the cores, one interesting detail in Fig.~\ref{fig:plot_VLOS_2_ZOOM} is the presence of a
filament that is feeding the central core. Although this was clearly identifiable in the 3D model due
to its different LOS velocity, it almost completely disappeared in synthetic observations, in the
line-velocity maps. This could be attributed to its much lower kinetic (and excitation) temperature.
The filament was detected in channel maps (for example in HCO$^{+}$), but only marginally. This is
because the width of the structure is below the beam size and the structure is weak compared to other
emission along the LOS. The detectability of such kinematic features should be studied further,
especially with models that more accurately describe the temperature and chemical composition of such
structures. 
 
The fact that radial velocities measured from synthetic ngVLA observations are close to those in ideal
observations will be important for studies of the accretion associated with cores and especially in the
context of high-mass star formation and hub-filament systems. Figure~\ref{fig:H13COp_velocity} shows
that ${\rm H^{13}CO^{+}}$(1-0) provides accurate measurements of the radial velocity and velocity
gradients, as judged by the comparison with ideal observations. 
This will be important, when gradients are used to detect infall and to estimate infall rates. For
example, \citet{Zhou2022_hubs} presented ALMA observations of hub-filament systems in a number of
proto-clusters. The systematic velocity gradients scales of a few $0.1$\,pc were up to tens of
km\,s$^{-1}$, suggesting almost free-fall motions towards the massive hubs. In
Fig.~\ref{fig:H13COp_velocity}, 0.1\,pc would correspond to $\sim 5\arcsec$, but the most systematic
velocity gradients are seen only close to the resolution limit, in agreement with the lower mass of
this system. 
However, in such complex regions, the estimation of velocity gradients and the corresponding gas-inflow
rates is complicated not only by the random inclinations of the structures but also by the
presence of many LOS structures that can lead to high apparent and partly artificial gradients. For
example, in Fig.~\ref{fig:H13COp_velocity}, the blue path crosses from diffuse background to a
filament, leading to an apparent gradient of close to $\sim50$\,km\,s$^{-1}$\,pc$^{-1}$ (around the
offset of $3 \arcsec$). Gradients should clearly be measured only along well-defined and continuous
structures. The risk for artificial gradients, caused by change alignment of LOS structures, may also
increase towards the central regions of hub-filament systems.

In Sect.~\ref{sect:PCA}, we looked briefly at the global velocity statistics with the help of the PCA
method. The PCA estimates of the $\delta v \sim R^{\alpha}$ scaling relation were well defined and the
slopes were almost identical for the $^{13}$CO and ${\rm C^{18}O}$ data, with $\alpha=0.78$. The
synthetic observations resulted, however, in clearly steeper relationships ($\alpha=1.00$ and
$\alpha=1.05$ for $^{13}$CO and ${\rm C^{18}O}$, respectively), again suggesting some effect from the
interferometric filtering. Of course, for such analysis of large-scale density or velocity fields, the
coverage of low spatial frequencies with additional single-dish data becomes crucial
(Sect.~\ref{sect:singleDish}). In spite of the S/N differences and the high S/N ${\rm C^{18}O}$
observations being limited just to the central filament, the $^{13}$CO and ${\rm C^{18}O}$ values were
quite consistent with each other.

\subsection{Failure of infall indicators} \label{disc:infall}

We compared the collapse indicator $\delta v$ of Eq.~(\ref{eq:infall_dv}) to the actual LOS flows in
the model. The latter were quantified with the parameter $\xi$, the density-weighted net mass flow
towards the highest LOS density peak (Eq.~(\ref{eq:infall_xi})). The comparison of $\xi$ and $\delta v$
showed little correlation (figures \ref{fig:infall_index_zoom} and \ref{fig:LOS_collapse_LOC},
respectively). This was true even in the case of ideal line observations.

The parameter $\delta v$ is based on the picture of a single core, where the excitation temperature
increases towards the centre. This leads to an asymmetry, where the optically thick species suffer more
self-absorption on its red-shifted side, leading to positive values of $\delta v$ (with the signs used
in Eq.~(\ref{eq:infall_dv})). This is clearly not an appropriate description for more complex regions,
where the LOS crosses several density peaks. In this case, $\delta v$ depends more on the random
superposition of emission from structures with different excitation temperatures and radial velocities.
Already in the case of two LOS clumps, the sign of $\delta v$ is likely to depend more on their
relative velocities than the infall within individual clumps. The same complexity also affects the
parameter $\xi$, but that at least is based on the true velocity field and concentrating on the main
LOS density peak.

Appendix~\ref{app:LOS_emission} illustrates the complexity of the density and velocity fields for
selected sightlines, where large $\xi$ values indicate a particularly strong inflow motion. We examine
here the LOS for the fourth core (counted from north) that is indicated in
Fig.~\ref{fig:collapse_spectra}. Figure~\ref{fig:plot_LOS_spectra_2} shows, how an observed {\rm
HCO$^{+}$(1-0) spectrum is composed of the emission and absorption at different locations along the
full LOS through the model cloud. In the case of the optically thinner ${\rm H^{13}CO^{+}}$, up to 30\%
of the local emission is absorbed by foreground layers, but the shape of the total spectrum is only
little affected by these optical-depth effects. In contrast, HCO$^{+}$ emission can be completely
absorbed by foreground structures, and most of the observed emission originates on the front side of
cloud. For individual cores, one can often observe the change from redshifted emission on the near side
to more blueshifted emission on the far side. However, several LOS structures contribute to the
emission and absorption at any given radial velocity, and the spatial separation of the structures is
$>$0.1\,pc, much larger than the size of individual cores. Because of the random radial-velocity
offsets between the structures, the observed spectrum does not show clear blue asymmetry. 

Figures in Appendix~\ref{app:LOS_emission} show further plots on how different LOS structures
contribute to different spectral lines. In particular, Fig.~\ref{fig:LOS_spectra_2} splits a spectrum
to its blue-shifted and red-shifted components in the case of another LOS. Also in that case, the
dominant core is consistent with the basic assumptions of the $\delta v$ diagnostic (red-shifted
emission on the front side and blue-shifted emission on the far side of the core), but this signature
is masked by other LOS structures. One similar concrete example was provided by \citet{Zhou2021}, who
carried out ALMA observations of the high-mass star-forming clump G286.21+0.17. The clump contains a
filament that, based on the blue asymmetry of the single-dish HCO$^{+}$ spectra, had been interpreted
to be in global collapse. Interferometric observations revealed two separate clumps, and the asymmetry
could be seen to be caused by the relative velocity of the clumps rather than by any systematic infall.

If all LOS structures were at clearly different radial velocities, the $\delta v$ statistics might be
applied to individual velocity components. However, it is difficult ascertain from observations that
one has correctly isolated a single core in velocity space, especially if the object is not fully
spatially resolved. Therefore, the $\delta v$-diagnostic will remain more applicable to nearby regions
of low-mass star formation, where the LOS confusion is lower and the spatial resolution typically
higher.

\begin{figure}
\centering
\sidecaption
\includegraphics[width=8.8cm]{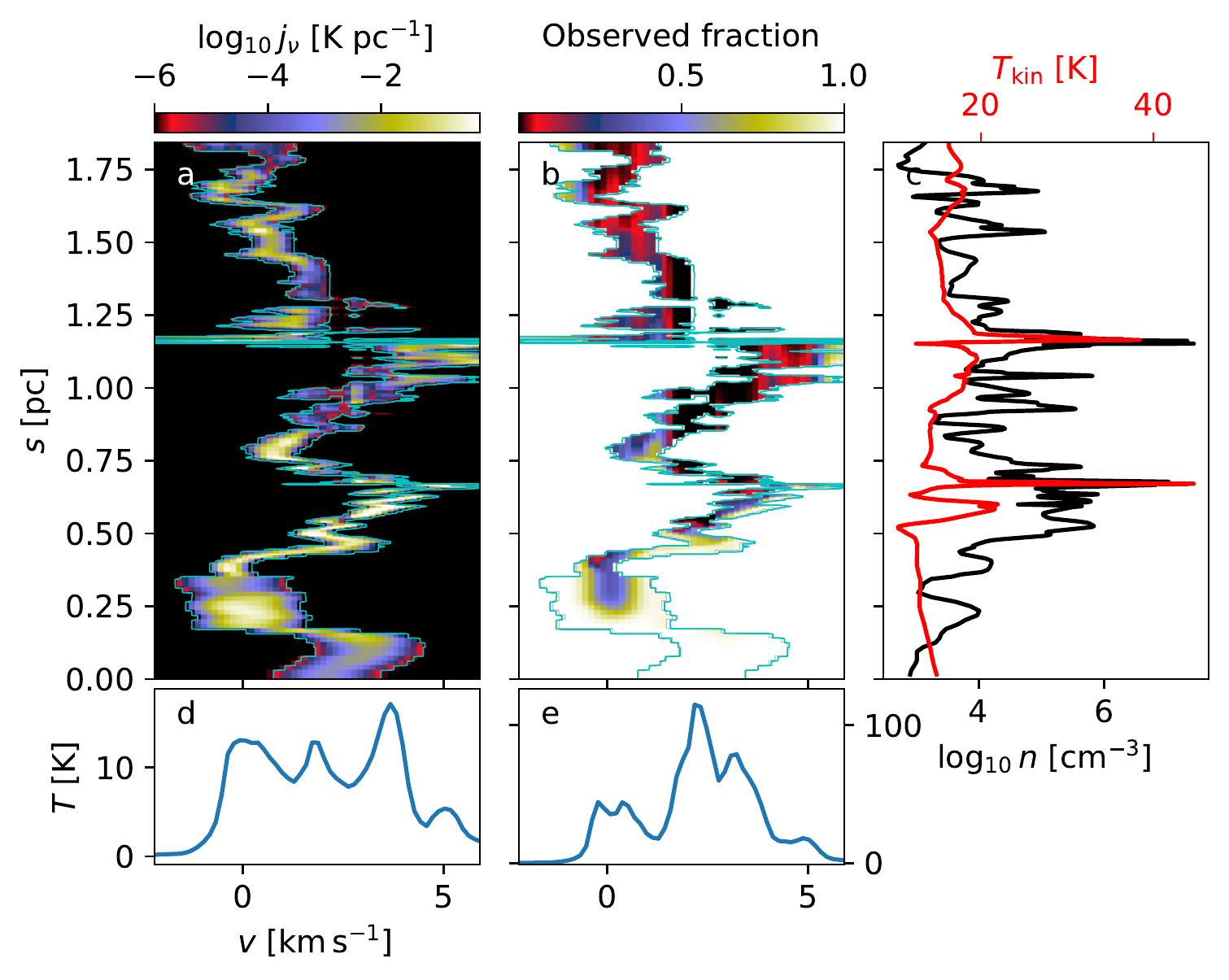}
\caption{
Contributions of different LOS regions to a selected ${\rm HCO^{+}}$(1-0) spectrum. In frames a-b,
the y-axis corresponds to the distance from the observer's side of the model cloud, and the x-axis
shows the radial velocity. Frame a shows the intensity of the local emission that reaches the observer
and thus contributes to the observed spectrum in frame d. Frame b shows the same as the fraction of the
local emission that is not absorbed by the foregrounds.
The cyan contours are drawn around significant emission, to help the comparison between the frames a
and b. In frame b, the observed fraction of emission is plotted only inside the contour.
Frame e shows the corresponding fictitious optically thin spectrum, where all foreground absorption is
ignored. The LOS variations in density (black curve) and kinetic temperature (red curve) as shown in
frame c.
}
\label{fig:plot_LOS_spectra_2}
\end{figure}

\subsection{Combination of interferometer and single-dish data \label{sect:singleDish} }

Interferometric measurements are usually combined with single-dish data to recover information of large
spatial scales \citep{Cornwell1993}. We simulated single-dish N$_{\rm 2}$H$^{\rm +}$(1-0) observations
for a 8.1$\arcsec$ beam, which corresponds to a 100-m dish such as the Green Bank Telescope
(GBT)\footnote{https://greenbankobservatory.org/science/telescopes/gbt/}. The observational noise was
set to 0.1\,K, which is realistic for GBT data \citep{Pingel2021}. The single-dish observations were
used as a model in the CASA tclean\footnote{https://casa.nrao.edu/docs/taskref/tclean-task.html}
procedure, resulting in a spectral cube with information also of the largest scales. As an alternative,
we reduced the ngVLA data separately and joined them with the single-dish data with the feathering
procedure. These two alternatives should produce comparable results, although in practice differences
up to tens of per cent might be observed in individual spectra \citep{Barnes2021}. We used the
uvcombine\footnote{https://github.com/radio-astro-tools/uvcombine} tool for the feathering of the ngVLA
ans single-dish images.

Figure~\ref{fig:SD_comparison} shows the results in terms of column densities and sample spectra at the
0.55$\arcsec$ resolution. The results are very similar, whether the ngVLA and single-dish data are
reduced together or joined by feathering. The spectral profiles are identical to within $\sim$10\%, and
the column densities along the selected stripe are also well correlated. Some larger differences are
observed close to the position $C$, where the wide lines and the presence of multiple velocity
components causes problems for our analysis that is based on hyperfine fits with a single velocity
component. With the exception of the position C, all observations tend to underestimate the true column
density. This in spite of the fact that the latter values are here again corrected for the spatial
abundance variations. Excluding the peak $N({\rm N_2 H^{+}})>2\times 10^{14}$\,cm$^{-2}$ around the
position $C$, the differences between the two alternative combinations of ngVLA and single-dish data
were $\sim$14\%, the feathering resulting in only 5\% lower average value.
Figure~\ref{fig:SD_comparison} also shows that the difference between the synthetic and the ideal ${\rm
N_2 H^{+}}$ observations is smaller than the difference between the ideal observations and the true
column densities. This applies both to the mean value and the variations along the selected stripe.

In Sect.~\ref{sect:colden}, column densities were calculated using single lines, because the
interferometric filtering appeared to have different effect on, for example, the optically thin ${\rm
C^{18}O}$ and the much more extended $^{13}$CO emission. After combining ngVLA and single-dish data,
Figure~\ref{fig:SD_others} shows correlations between the true column density (obtained directly from
the model) and estimates calculated using the ${\rm C^{18}O}$ vs. $^{13}$CO and the ${\rm
H^{13}CO^{+}}$ vs. HCO$^{+}$ line ratios. The situation is much improved over the ngVLA-only results in
Fig.~\ref{fig:plot_colden_cor}, but the use of the $^{13}$CO vs. ${\rm C^{18}O}$ line ratios results in
only marginal improvement over the simpler assumption of optically thin ${\rm C^{18}O}$ emission. There
is also some bias that is still larger in the synthetic observations than in the ideal observations.
Rather than having real physical causes, this might thus result from some imperfections in the data
reduction or in the inclusion of the single-dish data. Apart from systematic errors, the dispersion is
small and mostly below 25\%. The plot is limited to within 10$\arcsec$ of the two pointing centres,
where the hydrogen column densities are $N({\rm H_2})\sim 10^{23}$\,cm$^{-2}$ or even higher. The
estimates calculated from the ${\rm H^{13}CO^{+}}$ vs. HCO$^{+}$ line ratio do not show bias, but the
increase in dispersion towards lower column densities is more noticeable. 

In Fig.~\ref{fig:SD_others}a, the assumed $T_{\rm ex}=15$\,K resulted in quite accurate estimates. With
$T_{\rm ex}=25$\,K, the column densities would be overestimated by more than 50\%, the effect being
almost the same for the whole map. Thus, a single optically thin line can already provide much
information on relative structures, and, depending on the science case, the addition of good-quality
single-dish data can be more important than the more precise excitation-temperature information
obtained from line ratios.

\begin{figure}
\centering
\sidecaption
\includegraphics[width=8.8cm]{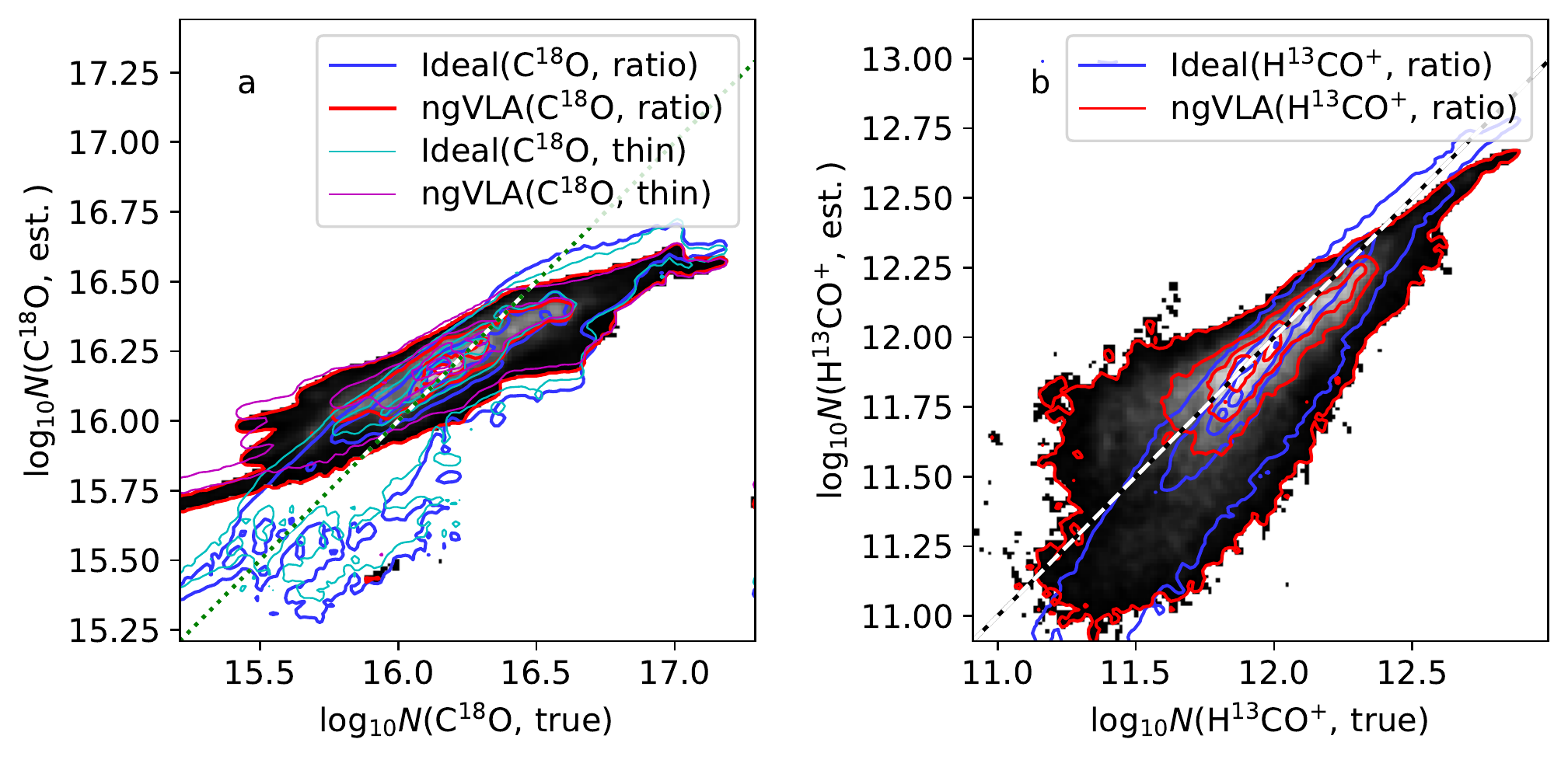}
\caption{ 
Column-density estimated from ${\rm C^{18}O}$ vs. $^{13}$CO (frame a) and ${\rm H^{13}CO^{+}}$ vs.
HCO$^{+}$ (frame b) line ratios, plotted against the true column density $N^{\prime}({\rm H}_2)$
(accounting for the effect of density-dependent fractional abundances). Results are shown for both
ideal observations and the combined ngVLA and single-dish data. The diagonal dashed line corresponds to
the expected one-to-one relation. For each dataset (listed in the legends) there are four contour
levels that are equidistant from close to zero up to 80\% of the maximum point density. The background
greyscale images correspond to the estimates calculated from ngVLA line ratios. In frame a, the
estimates for optically thin ${\rm C^{18}O}$ emission at $T_{\rm ex}$=15\,K are shown for comparison.
Plots use data within 10$\arcsec$ of the pointing centres and a resolution of 0.7$\arcsec$.
}
\label{fig:SD_others}
\end{figure}

The emission at the brightest parts of the filament was captured quite well already in the ngVLA-only
observations. Section~\ref{sect:velocity_fields} showed that the same conclusion applies to the
velocity fields observed towards the densest sub-structures. Therefore, ngVLA data without any
additional single-dish data could be sufficient for studies of the individual densest regions. On the
other hand, single-dish data remain essential for extended sources -- such as the examined IRDC model
-- and any studies into the large-scale velocity or density statistics (Sect.~\ref{sect:colden}).

\begin{figure}
\centering
\sidecaption
\includegraphics[width=8.8cm]{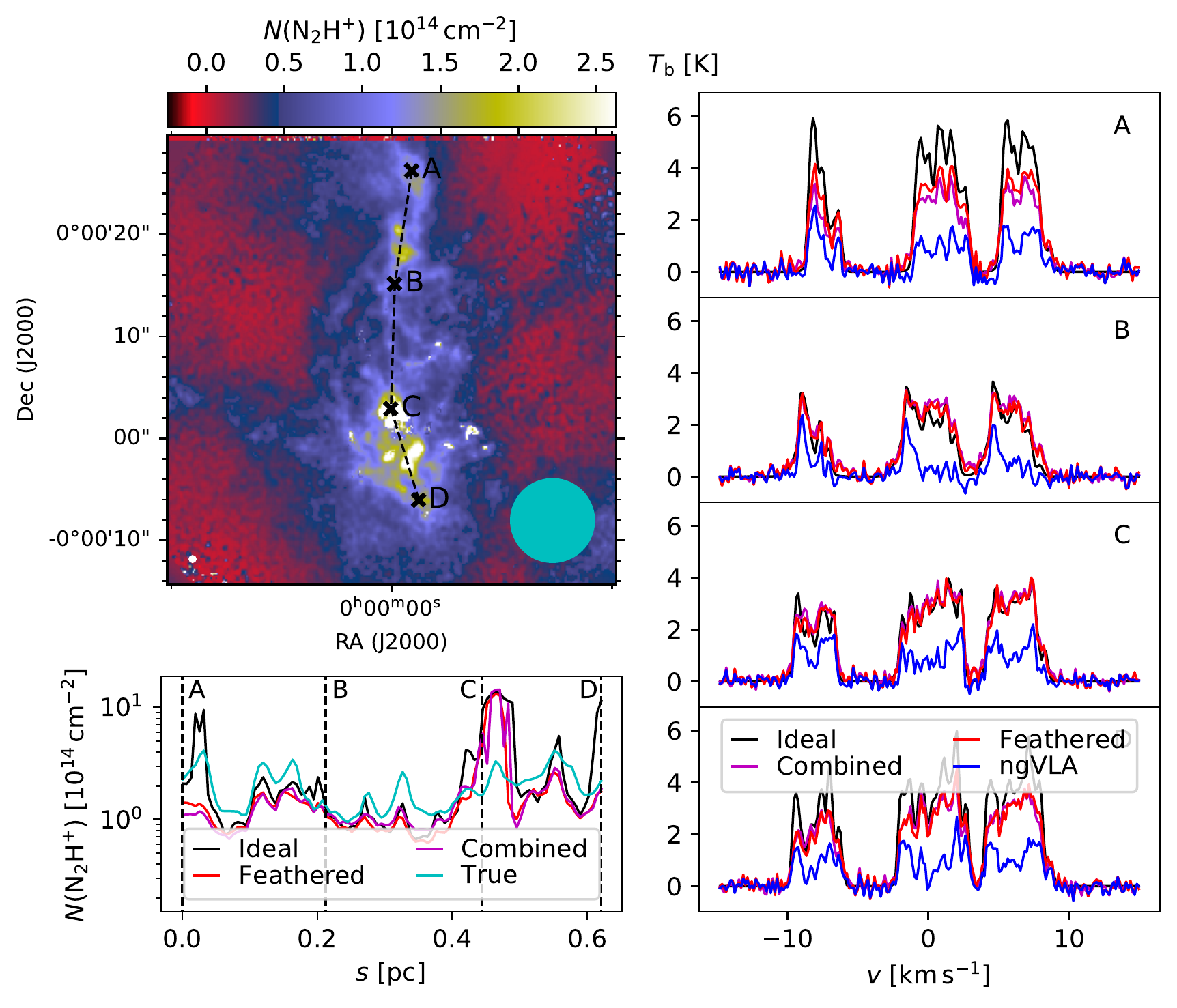}
\caption{
Comparison of ${\rm N_2 H^{+}}$ observations with and without single-dish data.  The top left frame
shows the \textit{N}(N$_{\rm 2}$H$^{\rm +}$) column-density map that is derived from the combined ngVLA
and single-dish data.  The circle in the lower left corner corresponds to the 0.55$\arcsec$ resolution
of the final map, and the circle in the lower right corner shows the 8.1$\arcsec$ beam of the
single-dish data. The bottom left frame shows a comparison of the column-density estimates along the
path that is plotted in the first frame. Results are shown for ideal observations and for ngVLA and
single-dish observations when both are included in the cleaning process (``combined'') or when they are
joined afterwards through feathering (``feathered''). We plot also the true column density that is
obtained directly from the model cube. The right-hand frames show sample spectra towards the four
positions marked in the first frame. The spectra from interferometric observations without combined
single-dish data (``ngVLA'') are also shown.
}
\label{fig:SD_comparison}
\end{figure}

\section{Conclusions}  \label{sect:conclusions}

We have simulated observations of a high-column-density molecular cloud to study the foreseen
performance of the ngVLA interferometer for studies of the ISM and the early phases of the
star-formation process. As the baseline case, we examined observations with the ngVLA Core antenna
configuration and a total observing time of six hours. The comparison of the ngVLA synthetic
observations, the corresponding ideal line observations, and the actual 3D cloud model resulted in the
following conclusions.
\begin{enumerate}
\item 
The ngVLA observations described the general structure of the filamentary cloud accurately. In the
measurements of the integrated line intensity (e.g. HCO$^{+}$, ${\rm N_2 H^{+}}$, and ${\rm NH_3}$),
the noise does not become significant even close to the maximum angular resolution (down to
$\sim$0.5$\arcsec$).
\item 
At high frequencies, observations of our model cloud suffer from some loss of extended emission. For
${\rm NH_3}$ the effect remains small at the scale of the examined cloud ($\sim 15\arcsec \times
40\arcsec$), and the column-density estimates remained accurate to about 30\%, even without the use of
single-dish data.
\item 
The kinetic temperatures derived from ${\rm NH_3}$ observations were mostly accurate to within
$\sim$1\,K. However, at lower column densities some positive bias was observed, which could be
attributed to some loss of extended signal even at the ${\rm NH_3}$(1,1) frequency.
\item 
The synthetic ngVLA observations provided an accurate image of the kinematics at intermediate scales.
For example, velocity gradients associated with the main cores could be traced mostly with a precision
better than $\sim$0.3\,km\,s$^{-1}$. At higher frequencies (e.g., ${\rm C^{18}O}$ and ${\rm N_2 H^{+}}$
lines), the radial-velocity data show low-spatial-frequency deviations, because the target cloud is
larger than the maximum recoverable scale. The loss of low-spatial-scale information is reflected in
global statistics, but the effects remained moderate, for example, in the PCA analysis of $^{13}$CO
(in spite of very extended emission) and ${\rm C^{18}O}$ (in spite of much lower
signal-to-noise ratio) lines.
\item 
The ngVLA observations traced the kinematics within the cores down to the resolution limit. However, at
the smallest scales some important features remain undetectable either because of beam dilution or
because their spectral signatures (even in ideal observations) are weak. The emission from some dense
regions was masked by the emission and absorption of other LOS regions, either because of temperature
differences or radiative transfer effects.
\item 
We compared standard collapse indicators to the actual infall motions in the 3D model cloud. The blue
asymmetry of optically thick lines was not significantly correlated with actual LOS motions in the
cloud. The spectra were complex, containing contributions from many LOS density peaks. This makes it
difficult to unambiguously interpret any observed spectral asymmetries in terms of a local collapse.
\item
The addition of single-dish data recovers the lost large-scale emission, and, for example, the
synthetic ${\rm N_2 H^{+}}$ spectra were found to be very similar to the ideal observations. However,
there can still be significant differences between the true column densities and the estimates, even if
these were derived from ideal observations. The complex velocity structure can lead to large errors or
even complete failure in the column-density estimation.
\end{enumerate}

\begin{acknowledgements}
EM is funded by the University of Helsinki doctoral school in particle physics and astrophysics (PAPU).
Tie Liu acknowledges the supports by National Natural Science Foundation of China (NSFC) through grants
No.12073061 and No.12122307, the international partnership program of Chinese Academy of Sciences
through grant No.114231KYSB20200009, Shanghai Pujiang Program 20PJ1415500 and the science research
grants from the China Manned Space Project with no. CMS-CSST-2021-B06.
We thank Troels {Haugb{\o}lle} for providing the data for the MHD simulations, and we acknowledge PRACE
for awarding access to Curie at GENCI@CEA, France to carry out those simulations. 
\end{acknowledgements}

\bibliographystyle{aa}
\bibliography{my.bib}

\appendix

\section{Maps of dust emission and temperature} \label{sect:Tdust}

The line simulations assumed that the kinetic temperature is equal to the dust
temperature (Sect.~\ref{sect:simulations}). Figure ~\ref{fig:plot_Tdust} shows for
reference maps of the 250\,$\mu$m surface brightness and dust colour temperature. The
temperatures are estimated with modified blackbody fits to the synthetic 160, 250, 350,
and 500\,$\mu$m surface brightness maps, giving equal weight to data in all four bands.
The dust opacity spectral index was assumed to be a constant $\beta$=1.8, which is close
to the value in the used dust model.

\begin{figure}
\sidecaption
\centering
\includegraphics[width=8.8cm]{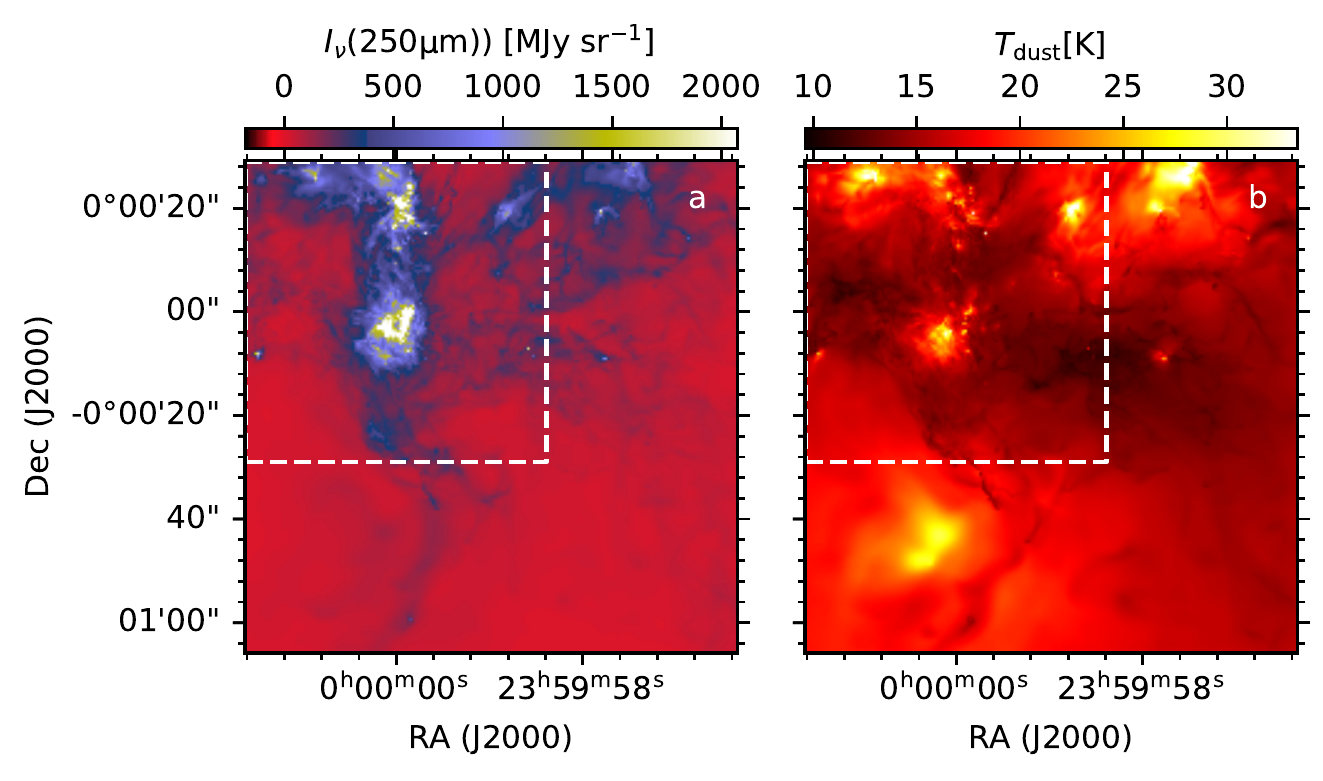}
\caption{
Maps of dust emission at 250\,$\mu$m (left frame) and dust colour temperature (right
frame). The dashed line shows the area discussed in the main part of the paper. The data
correspond to the full resolution of the cloud model (using 0.05$\arcsec$ pixel size for
the plot).
}
\label{fig:plot_Tdust}
\end{figure}

\section{Origin of observed emission} \label{app:LOS_emission}

Figure~\ref{fig:PPP_PPV} shows the main structures of the 3D density field and the
kinematic structures seen in the PPV cube of the synthetic $^{13}$CO data. The density
values were thresholded at $n=5\times 10^5$\,cm$^{-3}$ and the detected regions, each
with an individual label, were further extended to the region with $n>3\times
10^5$\,cm$^{-3}$. Regions with fewer than 100 cells were removed, where the cell size
corresponded to 0.25$\arcsec$. The $^{13}$CO data were analysed using the results from
Gaussian fits with up to three velocity components, only using the components with
brightness temperatures exceeding 3.5 K. These provide discrete points in the PPV space,
where neighbouring points were further connected, if their distance was below
\begin{equation}
  \sqrt{ \left(\frac{\Delta x}{\delta x}\right)^2 + \left(\frac{\Delta v}{\delta v}\right)^2} < 2.
\end{equation}
Here $\Delta v$ and $\Delta x$ are the distances along the velocity and spatial
coordinates. The scales were set to $\delta v = 0.2$\,km\,s$^{-1}$ and to $\delta x$
equal to half of the beam FWHM. Each connected PPV region (velocity-coherent region) was
given a unique label. Figure \ref{fig:PPP_PPV} shows the extracted PPP and PPV regions as
stereographic images.

\begin{figure*}
\centering
\sidecaption
\includegraphics[width=12cm]{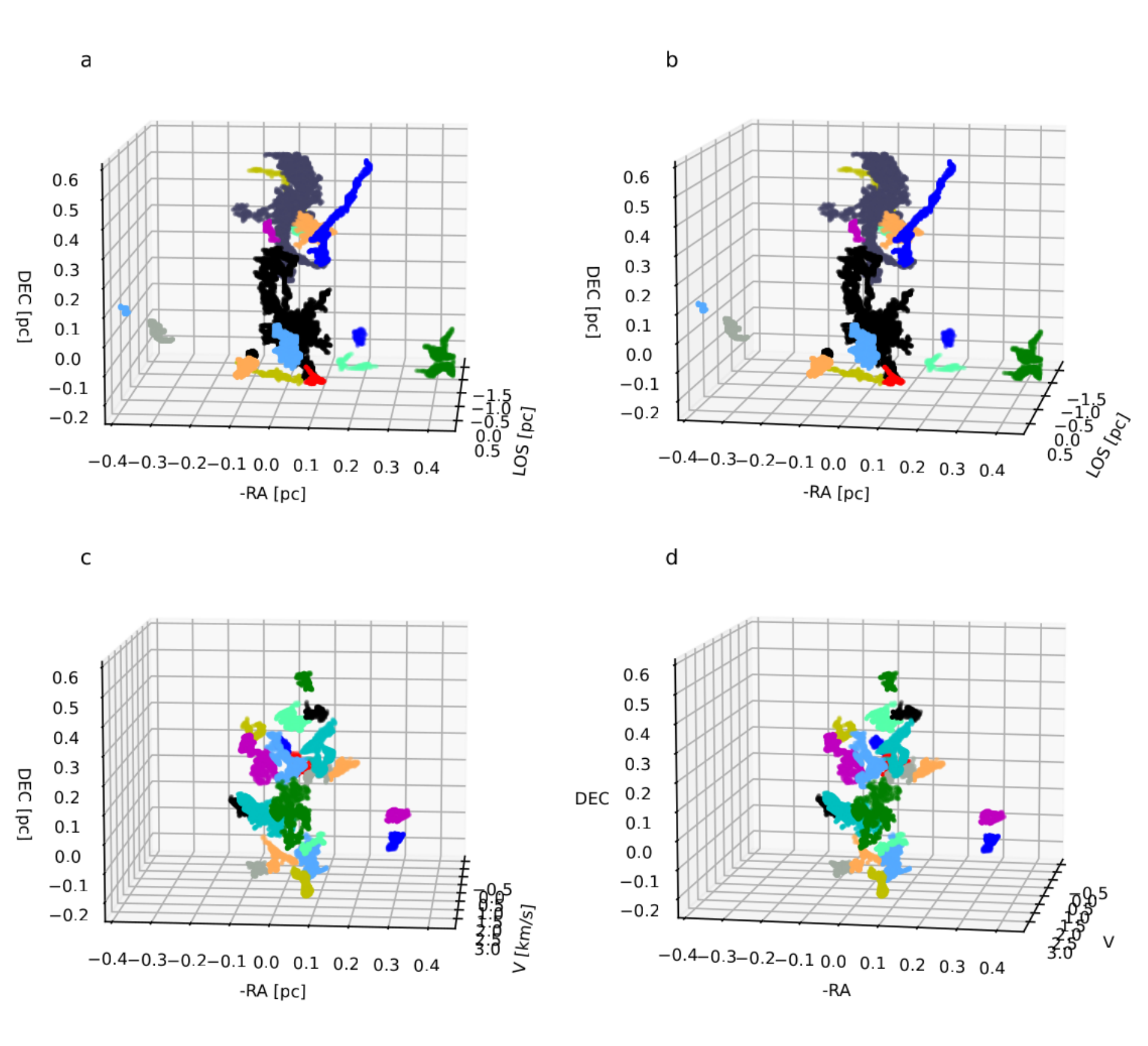}
\caption{
Stereographic images of the high-density structures (frames a and b) and the PPV
structures extracted from synthetic $^{13}$CO ngVLA observations (frames c and d). The
density structures correspond to regions above $n=3\times 10^5$cm$^{-3}$ and the PPV
structures to individual Gaussian components brighter than 3.5\,K in brightness
temperature.
}
\label{fig:PPP_PPV}
\end{figure*}

Because the usual collapse indicator $\delta v$ showed little correlation with the actual
infall motions in the model cloud, we examined further the contribution of different LOS
regions to the observed spectra. Figure~\ref{fig:collapse_spectra} shows spectra towards
five positions with large $\xi$ values. These all contain several velocity components.

\begin{figure*}
\centering
\sidecaption
\includegraphics[width=11cm]{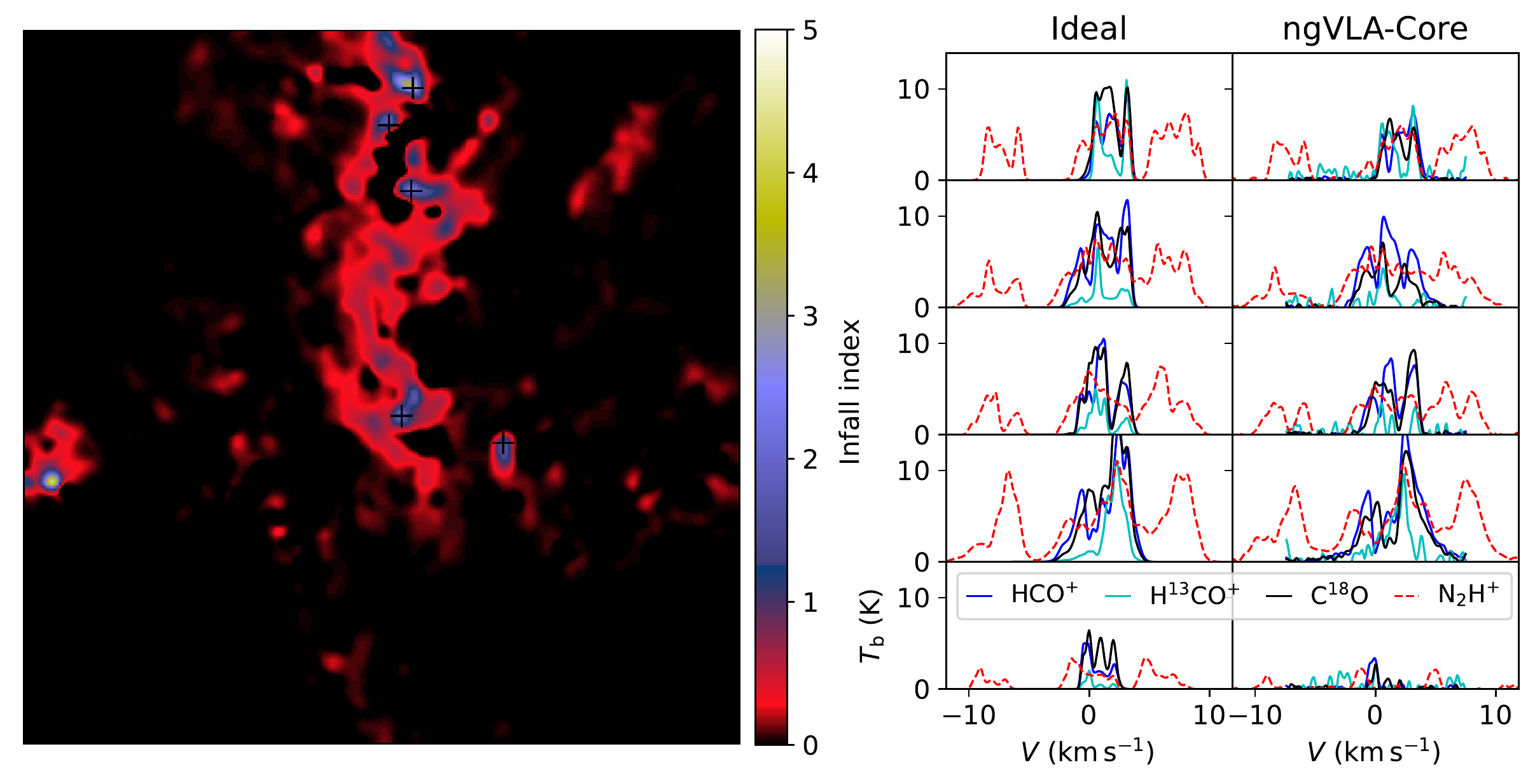}
\caption{
Spectra for selected LOS with strong inflow motions. The first frame shows the inflow
index $\xi$ convolved to 1$\arcsec$ resolution. Crosses mark five locations with large
$\xi$ values , and the plots on the right show spectra for these positions (rows in
decreasing order of latitude). Spectra in the first column correspond to ideal
observations and the spectra in the second column to simulated ngVLA observations.
Spectra are plotted for ${\rm C^{18}O}$ (black lines), HCO$^{+}$ (blue), ${\rm
H^{13}CO^{+}}$ (cyan), and ${\rm N_2 H^{+}}$ (dashed red lines).
}
\label{fig:collapse_spectra}
\end{figure*}

We examine further the northernmost LOS marked with a cross in
Fig.~\ref{fig:collapse_spectra}. Figure~\ref{fig:LOS_spectra_1} plots the density,
kinetic temperature, and relative abundance along the full LOS. The LOS consist of a
number of smaller density peaks, some of which are close in density to the strongest one.
The lower frames of Fig.~\ref{fig:collapse_spectra} show the contributions of different
LOS regions to the intensity in the observed spectrum. One takes here into account not
only the local emission but also how the intensity is attenuated by foreground
absorption. There are several density peaks with almost equal contributions to the
observed spectra.

In Fig.-\ref{fig:LOS_spectra_2}, we examine further two pairs of spectra, ${\rm C^{18}O}$
and $^{13}$CO and, on the other hand, ${\rm H^{13}CO^{+}}$ and HCO$^{+}$. We separate the
lines to their red-shifted and blue-shifted parts using the mean velocity of the
optically thinner line.
The main density peaks (indicated with arrows) are associated with red-shifted emission
at the front side and blue-shifted emission at the back side, which is consistent with
collapse motions and agrees with assumptions of the $\delta v$ diagnostic. However, in
the final spectra such local effects are masked by the superposition of emission from
many density peaks. This supports a scenario where, even if there were systematic inflow
at all scales, the values of $\delta v$ will depend more on the relative density,
excitation, and radial velocity of individual density structures, rather than the
emission asymmetry in individual cores.

\begin{figure*}
\centering
\sidecaption
\includegraphics[width=11cm]{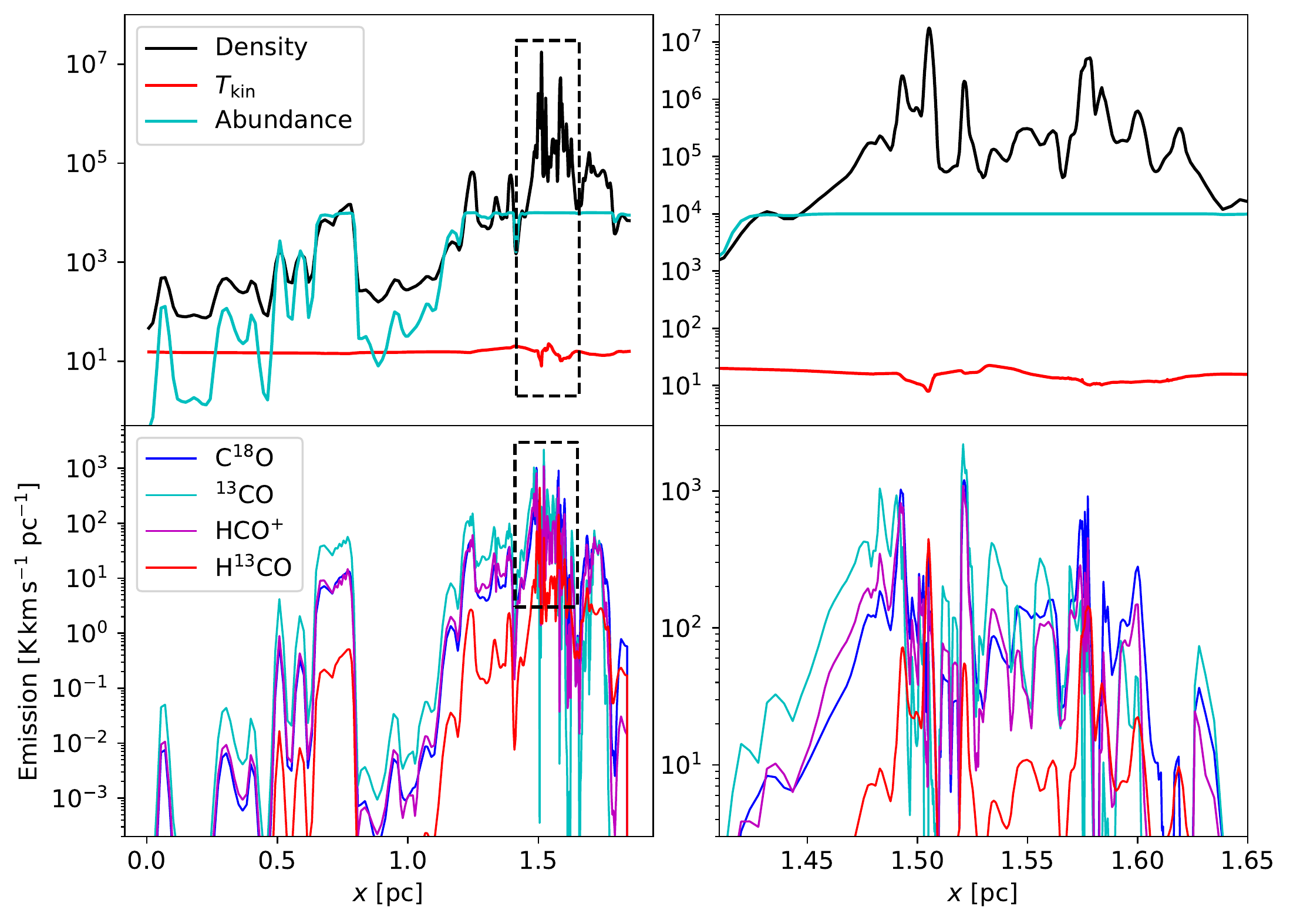}
\caption{
Parameters along the LOS towards the position indicated by the northernmost cross in
Fig.~\ref{fig:collapse_spectra}. The first frame shows the density (black line), abundance (arbitrary
normalisation and cyan line), and kinetic temperature (red line). Frame b shows same data with a
zoom-in to the dashed box in frame a. The bottom frames shows the contributions of different LOS
regions to the observed spectra (integrated line intensity).
}
\label{fig:LOS_spectra_1}
\end{figure*}

\begin{figure*}
\centering
\sidecaption
\includegraphics[width=11cm]{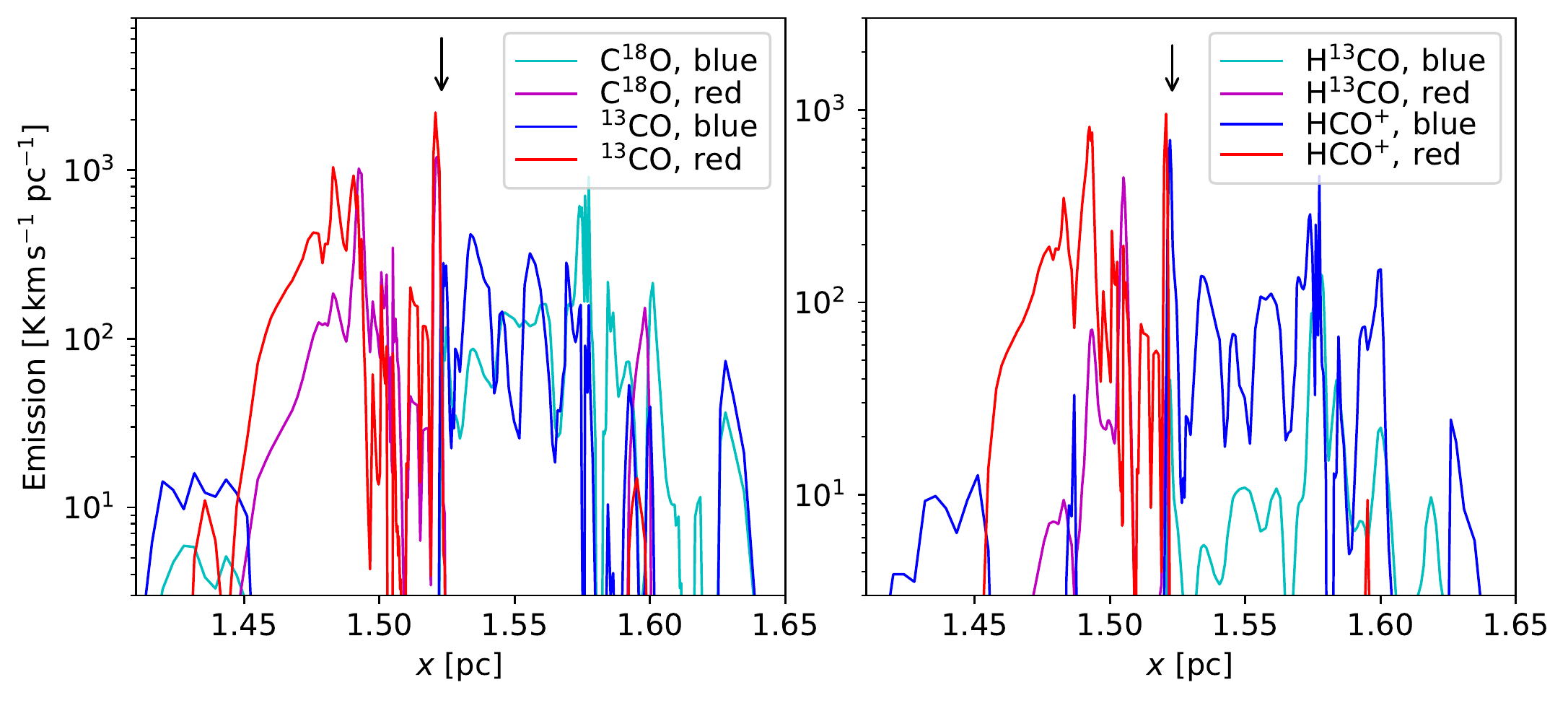}
\caption{
Same as lower right frame of Fig.~\ref{fig:LOS_spectra_1} but splitting emission to blue-shifted and
red-shifted parts at the mean velocity of the optically thinner species (${\rm C^{18}O}$ in the left
and ${\rm H^{13}CO^{+}}$ in the right frame). The black arrows indicate the position of the main
density peak.
}
\label{fig:LOS_spectra_2}
\end{figure*}

\section{Column density estimates} \label{app:colden}

The observed brightness temperatures are 
\begin{equation}
T_{\rm b} =  \eta \, [ J_{\nu}(T_{\rm ex})-J_{\nu}(T_{\rm bg}) ] \, (1-e^{-\tau}),
\end{equation}
where we assume an efficiency (including beam filling) of $\eta=1$ and a background temperature of
$T_{\rm bg}$=2.73\,K (same as in the radiative transfer simulations), and function $J$ is defined as
\begin{equation}
J_{\nu}(T) = \frac{h \nu / k}{e^{h \nu / k T}-1}.
\end{equation}

For the ${\rm NH_3}$(1,1) and ${\rm N_2 H^{+}}$ lines the optical depth $\tau$ and the
excitation temperature $T_{\rm ex}$ are obtained directly from the fitting of the above
equation to the observed spectra, simultaneously including all hyperfine components with
their known relative intensities.  

Optical depth can also be estimated from the intensity ratio of lines with a known
opacity ratio. This is often used with ${\rm C^{18}O}$ and $^{13}$CO lines, and we apply
the method to the HCO$^{+}$ and ${\rm H^{13}CO^{+}}$ lines. Assumptions of the same
excitation temperature, beam filling, and line width for both lines lead to the
expression
\begin{equation}
\frac{T_{\rm b}(A)}{T_{\rm b}(B)} = \frac{1-e^{-r \times \tau(B)}}{1-e^{-\tau(B)}}
\end{equation}
for the ratio of the measured brightness temperatures \citep{Myers1983}. With the assumed
ratio $r$ between the optical depths of the two species $A$ and $B$, the optical depths
can be solved.

Once the optical depth $\tau_{\nu}$ and excitation temperature $T_{\rm ex}$ are known, the
column density of molecules on the upper energy level of the transition can be calculated
as
\begin{equation}
N_u = \frac{8 \pi \nu^3}{c^3 A_{u l}} \frac{\int \tau_{\nu} d v}{e^{\frac{h \nu}{k T_{\rm ex}}} - 1},
\end{equation}
or, using the peak optical depth $\tau_0$ and FWHM line width $\Delta v$ from Gaussian fits,
\begin{equation}
 N_u = \frac{4 \pi^{3 / 2}}{\sqrt{\ln 2}}  
 \frac{\nu^3 \tau_{_0} \Delta  v}{c^3 A_{u l} 
 \left( e^{\frac{h \nu}{k T_{\rm ex}}} - 1 \right)} . 
\end{equation}
The total column density of the molecules is obtained by scaling with the ratio of the sum of
populations on all energy levels relative to the population of the level $u$,
\begin{equation}
\Gamma = \frac{Q}{g_u e^{-E_{\rm u}/kT_{\rm ex}}} = \frac{\sum_{i} g_i e^{-E_i/k T_{\rm ex}} }{g_u e^{-E_{\rm u}/kT_{\rm ex}}},
\end{equation}
assuming the same excitation temperature for all transitions. Here $g_i$ are the
statistical weights and $E_i$ the energies of the energy levels, and $Q$ is the partition
function.

If the emission is optically thin, column density estimates can be written in terms of an assumed
excitation temperature $T_{\rm ex}$ and the integrated line intensity $W$,
\begin{equation}
N_{\rm tot} = \frac{8 \pi \nu^3 W Q}{g_u c^2 A_{ul}} 
\frac{e^{E_l/k T_{\rm ex}}}{1-e^{-h \nu / k T_{\rm ex}}}
\frac{1}{J_{\nu}(T_{\rm ex})-J_{\nu}(T_{\rm bg})}.
\end{equation}
\citep{Caselli2002b}. We use this in calculations based on Gaussian fits, where
\begin{equation}
W = \sqrt{\frac{\pi}{4 \log 2}} \, T_{\rm b} \, \Delta v.
\end{equation}
In particular, for the $J = 1 \rightarrow 0$ transition of C$^{18}$O, Mangum \& Shirley
(2015) provide the formula
\begin{equation}
N_{\rm tot}  \left( {\rm C^{18}O} \right)
   =  \frac{2.48 \cdot 10^{14} (T_{\rm ex} + 0.88) e^{T_0 / T_{\rm ex}}  \int T_b dv {\rm (km/s)}}
   {\eta [e^{T_0 / T_{\rm ex}} - 1]  [J_{\nu} (T_{\rm ex}) - J_{\nu} (T_{\rm bg})]},
\end{equation}
where $T_0 = \frac{h \nu}{k} = 5.27$\,K.
Similarly, for optically thin ${\rm N_2H^+}$ emission
\begin{equation}
 N_{\rm tot} \left( {\rm N_{2}H^{+}} \right) = \frac{6.25 \cdot
   10^{15} e^{T_0 / T}  (T_b / R_i) \Delta v}{\nu \eta [e^{T_0 / T} - 1] 
   [J_{\nu} (T_{\rm ex}) - J_{\nu} (T_{\rm bg})]}  [{\rm cm}^{- 2}].
\end{equation}
In these formulas, frequency $\nu$ is given in units of GHz and the line width $\Delta v$ in units of km
s$^{- 1}$, and the result is in units of cm$^{-2}$. In the ${\rm N_2H^{+}}$ equation, $T_b$ refers
to the brightness temperature of one of the hyperfine components, which is then scaled with the
relative intensity $R_i$ of that component ($\Sigma_i R_i \equiv 1$).  However, if the line is not
optically thin, one can estimate $T_{\rm ex}$ and $\tau$ by a simultaneous fit to all hyperfine
components, as mentioned above.

The hyperfine fit to ammonia ${\rm NH}_3 (1, 1)$ spectra also provides $T_{\rm ex}$ and $\tau$.
Assuming Gaussian line parameters, the integration of optical depth profile gives
\begin{equation}
 N_u ({\rm NH}_3  (1, 1)) = 1.6 \cdot 10^{13}  \frac{\tau (1, 1, m)
   \Delta v}{e^{T 0 / T} - 1}  
\end{equation}
for the column density of the upper level, with $T_0 = 1.14$ K, the line width $\Delta v$ being
included in units of km s$^{- 1}$ \citep{Harju1993}. As in the case of ${\rm N_2H^{+}}$, if the
emission is too weak for the hyperfine fit, column density can still be estimated from the integrated
brightness temperature that is assumed to be optically thin,
\begin{equation}
 N_u = \frac{1}{\eta} \xi \frac{1}{1 - \frac{\exp (T_0 / T_{\rm ex}) -
   1}{\exp (T_0 / T_{{\rm bg}}) - 1}}  \int T_b d v \quad [{\rm cm}^{- 2}] .
\end{equation}
The NH$_3$(2, 2) transition can be handled in the same way, integrating only over the main component.
The numerical factor $\xi$ is $1.3 \cdot 10^{13}$ for the NH$_3$(1,1) and $6.2 \cdot 10^{12}$ for the
NH$_3$(2,2) line. The total column density of NH$_3$(1,1) is obtained as the sum of populations on
the upper and lower level,
\begin{equation}
 N ({\rm NH}_3 (1, 1)) = N_u (1 + e^{T_0 / T_{{\rm ex}}}).
\end{equation}
The rotation temperature $T_{12}$ is obtained from the column density ratio
\begin{equation}
 \frac{N (2, 2)}{N (1, 1)} = \frac{5}{3} e^{\Delta E / k T_{12}}.
\end{equation}
\citet{Walmsley_Ungerechts_1983} provide the transformation from $T_{12}$ to kinetic temperature, based
on a three-level model of the (1,1), (2,2), and (2,1) levels of para ammonia. Finally, the total
ammonia column density can be estimated as
\begin{equation}
 N ({\rm NH}_3) = N (1, 1)  \left\{ \frac{1}{3} e^{23.4 / T_{12}} + 1 +
   \frac{5}{3} e^{- 41.5 / T_{12}} + \frac{14}{3} e^{- 101.5 / T_{12}}
   \right\}, 
\end{equation}
assuming that only metastable levels are populated \citep{Ungerechts1986, Harju1993}.

\section{Fits of spectral profiles} \label{sect:spectral_fits}

Most of the analysis was based on Gaussian fits or, in the case of ${\rm N_2 H^{+}}$ and ${\rm NH_3}$,
simultaneous fitting of all hyperfine components and assuming a single velocity component. All Gaussian
fits were also repeated with up to three velocity components. The calculations were made by our own
GPU-accelerated fitting routine, with approximately 1\,ms run time per spectrum (less for
single-component Gaussians, slightly more in case of multiple velocity components or hyperfine
spectra). In Sect.~\ref{sect:singleDish}, the pyspeckit program \citep{pyspeckit} was used for
comparison and to confirm the correct performance of our own routine.

When the spectra contain multiple velocity components, a fit may converge to a wrong solution, failing
to fit the main velocity feature or, in fits with $N_{\rm C}$ velocity components, failing to fit the
$N_{\rm C}$ most important features. Each fit was therefore repeated four times, using different
initial parameter values and keeping the results from the fit with the lowest $\chi^2$ value.
Figure~\ref{fig:fit_c18o_CASA} shows examples of the fits of ${\rm C^{18}O}$ spectra towards the
northern core, using 1-3 velocity components. Not all fits are perfect (e.g., three-component fits
sometimes missing one of the three most important features, like in the second frame of the plot). On
the other hand, even the single-component fits usually approximate the total emission well, biasing the
column-density estimates by much less than 50\%.

\begin{figure}
\centering
\sidecaption
\includegraphics[width=8.8cm]{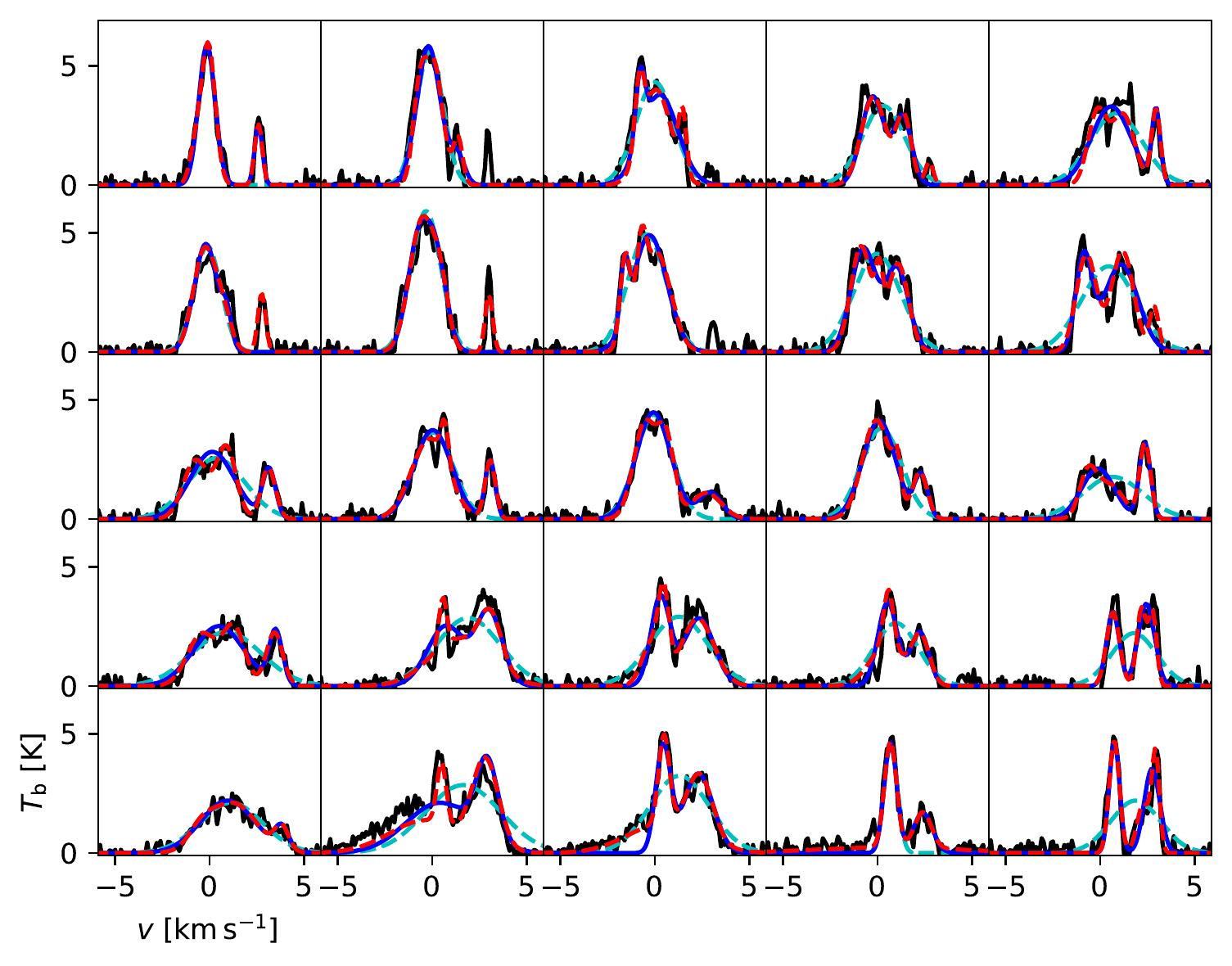}
\caption{
Examples of Gaussian fits towards the northern core, for simulated ${\rm C^{18}O}$
observations (black lines) with the ngVLA Core configuration.  The fits with one, two,
and three Gaussian components are plotted with cyan, blue, and red lines, respectively.
The spectra are taken at steps of FWHM, equal to 0.54$\arcsec$.
}
\label{fig:fit_c18o_CASA}
\end{figure}

The presence of multiple velocity components presents problems especially for hyperfine fits, because,
without additional constraints, some components might converge to an unphysical solution (e.g. a
combination of a very low $T_{\rm ex}$ and a very large optical depth). This is not possible in the
case of a single component that alone must match the observed spectrum. Conversely, if a spectrum
that contains multiple velocity components is fitted with a model that contains only a single velocity
component, the results are again biased. Figure~\ref{fig:test_HF_fit_MC2} shows, how the recovered
optical depth and column density depend on the velocity difference between two equally strong velocity
components. In this example, the original optical depth (the sum of the hyperfine components) is ten,
and the velocity difference is increased from zero to 4.0\,km\,s$^{-1}$. The two components both
have (before optical-depth effects) a line width of $\Delta v=$1.0\,km\,s$^{-1}$. As the velocity
components move further apart and the fitted spectrum falls between them, the optical depth is
increasingly underestimated. On the other hand, the estimated column density, which also depends on
both $T_{\rm ex}$ and $\Delta v$, remains remarkably accurate. 

For large velocity differences, $\delta v \ga 2$\,km\,s$^{-1}$, we occasionally saw alternative fits
that indicated much higher optical depths. These can be understood as an attempt to match the two
velocity components with a single wide, flat-topped (i.e. completely saturated) spectral profile. These
solutions tended to have higher $\chi^2$ values than the low-$\tau$ solutions, such as shown in
Fig.~\ref{fig:test_HF_fit_MC2}, suggesting that the alternative solutions might be caused by poor
convergence or a local $\chi^2$ minimum. The differences in the $\chi^2$ values between the good and
the alternative worse fit was only at 10\% level. Therefore, one might encounter more of such bad fits
(although possibly corresponding to a global $\chi^2$ minimum) when the signal-to-noise ratio of the
observations is lower.
They would then result in both the optical depth and the column density to be overestimated,
potentially even by a factor of a several.

\begin{figure}
\centering
\sidecaption
\includegraphics[width=8.8cm]{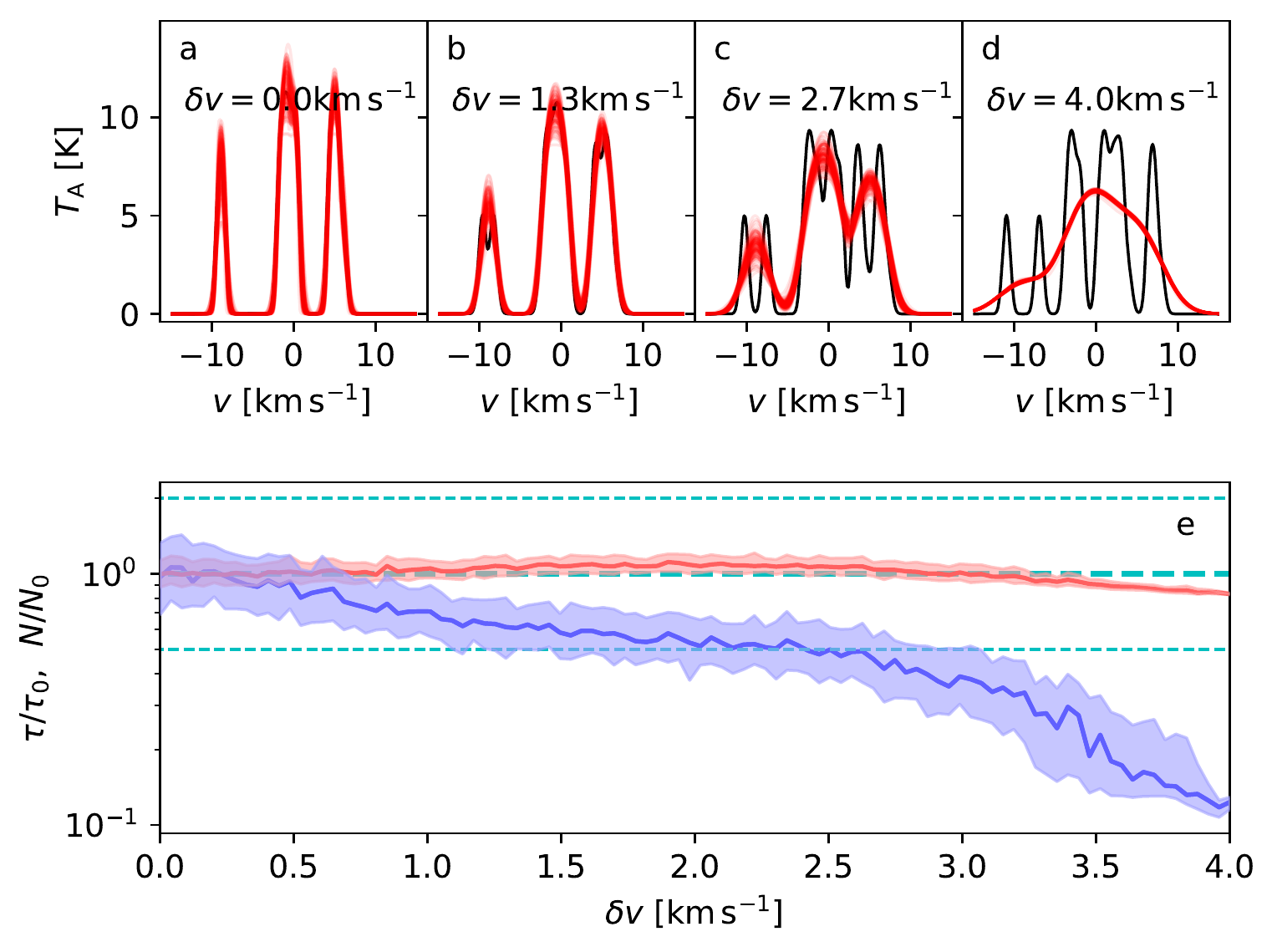}
\caption{
Test of fitting N$_2$H$^{+}$ observations containing two velocity components. The upper
frames show simulated spectra (black lines) with 0.3\,K observational noise and varying
velocity difference $\delta v$ between the two components. Fits of a single velocity
component are plotted in red for a number of noise realisations. The lower frame shows
the recovered optical depth (blue line) and column density (red line) as a function of
the velocity offset $\delta v$. The shaded regions correspond to the inter-quartile
range, which is computed from 128 noise realisations. The y-axis shows the ratio between
the estimates and the true values.
}
\label{fig:test_HF_fit_MC2}
\end{figure}

\end{document}